\documentclass[11pt]{article}
\usepackage{fullpage}
\usepackage{microtype}
\usepackage[OT1]{fontenc}
\usepackage{amsmath}
\usepackage{amssymb}
\usepackage{amsthm}
\IfFormatAtLeastTF{2026-06-01}{}{\usepackage{aliascnt}}
\usepackage[linesnumbered,boxed,ruled,vlined]{algorithm2e}
\usepackage{algpseudocode}
\usepackage{enumitem}
\usepackage{xifthen}
\usepackage{xspace}

\usepackage[margin=1cm]{caption} % Adds an addition margin on either side of figure captions. This must come before "\usepackage{subfig}" (otherwise there are errors).
\usepackage{subfig}

\usepackage[thinlines]{easytable}

\usepackage[bookmarks=true,hypertexnames=false,pagebackref]{hyperref}
\hypersetup{colorlinks=true, citecolor=blue, linkcolor=red, urlcolor=blue}

\usepackage{pgfplots}
\pgfplotsset{compat=1.16}
\usepackage{tikz}
\usetikzlibrary{arrows,arrows.meta,backgrounds,calc,fit,decorations.pathreplacing,decorations.markings,shapes.geometric}

\tikzstyle{internal} = [draw, fill, shape=circle]
\tikzstyle{external} = [shape=circle]
\tikzstyle{square}   = [draw, fill, rectangle]
\tikzstyle{triangle} = [draw, fill, regular polygon, regular polygon sides=3, inner sep=3pt]
\tikzstyle{pentagon} = [draw, fill, regular polygon, regular polygon sides=5, inner sep=2pt, minimum size=14pt]
\tikzset{every fit/.append style=text badly centered}

\usetikzlibrary{positioning,chains,fit,shapes,calc}
\usetikzlibrary{trees}
\usetikzlibrary{decorations.pathreplacing}
\usetikzlibrary{decorations.pathmorphing}
\usetikzlibrary{decorations.markings}
\tikzset{>=latex} % arrow tips
\usepackage{ifthen}

\usepackage{cleveref}

\usepackage[textsize=tiny]{todonotes}

\usepackage[normalem]{ulem}

\usepackage{mleftright}

\usepackage{cool}
\Style{DSymb={\mathrm d},DShorten=true,IntegrateDifferentialDSymb=\mathrm{d}}

\newcommand{\tp}[1]{{\left( #1 \right)}}

\def\*#1{\mathbf{#1}}
\def\+#1{\mathcal{#1}}
\def\-#1{\mathrm{#1}}
\def\=#1{\mathbb{#1}}
\def\^#1{\mathbb{#1}}

\newcommand{\norm}[2]{\ensuremath{\Vert #2 \Vert_{#1}}}
\newcommand{\abs}[1]{\ensuremath{\left\vert#1\right\vert}}
\newcommand{\inner}[2]{\ensuremath{\left\langle #2 \right\rangle}_{#1}}
\newcommand{\ceil}[1]{\lceil#1\rceil}
\newcommand{\floor}[1]{\lfloor#1\rfloor}

\newcommand{\mixingtime}[1]{\ensuremath{t_{\textnormal{mix}}(#1)}}

\newcommand{\dist}{\operatorname{dist}}

\newcommand{\transpose}[1]{\ensuremath{#1^{T}}}

\newcommand{\defeq}{:=}

\newcommand{\DTV}[2]{\-D_{\mathrm{TV}}\left({#1},{#2}\right)}

\IfFormatAtLeastTF{2026-06-01}{%
  \newcommand{\newsharedtheorem}[2]{\newtheorem{#1}[theorem]{#2}}%
}{%
  \newcommand{\newsharedtheorem}[2]{%
    \newaliascnt{#1}{theorem}%
    \newtheorem{#1}[#1]{#2}%
    \aliascntresetthe{#1}%
  }%
}

\newtheorem{theorem}{Theorem}
\newsharedtheorem{conjecture}{Conjecture}
\newsharedtheorem{lemma}{Lemma}
\newsharedtheorem{claim}{Claim}
\newsharedtheorem{observation}{Observation}
\newsharedtheorem{proposition}{Proposition}
\newsharedtheorem{corollary}{Corollary}
\theoremstyle{definition}
\newsharedtheorem{condition}{Condition}
\newsharedtheorem{definition}{Definition}

\theoremstyle{remark}
\newtheorem*{remark}{Remark}

\crefname{theorem}{Theorem}{Theorems}
\crefname{observation}{Observation}{Observations}
\crefname{claim}{Claim}{Claims}
\crefname{condition}{Condition}{Conditions}
\crefname{algorithm}{Algorithm}{Algorithms}
\crefname{property}{Property}{Properties}
\crefname{example}{Example}{Examples}
\crefname{fact}{Fact}{Facts}
\crefname{lemma}{Lemma}{Lemmas}
\crefname{corollary}{Corollary}{Corollaries}
\crefname{definition}{Definition}{Definitions}
\crefname{remark}{Remark}{Remarks}
\crefname{proposition}{Proposition}{Propositions}
\crefname{equation}{equation}{equations}
\crefname{enumi}{Case}{Case}
\creflabelformat{enumi}{(#2#1#3)}

\makeatletter
\def\prob#1#2#3{\goodbreak\begin{list}{}{\labelwidth\z@ \itemindent-\leftmargin
      \itemsep\z@  \topsep6\p@\@plus6\p@
      \let\makelabel\descriptionlabel}
  \item[\textbf{Name}]#1
  \item[\textbf{Instance}]#2
  \item[\textbf{Output}]#3
  \end{list}}
\makeatother

\makeatletter
\providecommand\@dotsep{5}
\def\listtodoname{Todo list}
\def\listoftodos{\@starttoc{tdo}\listtodoname}
\makeatother

\def\branden{Br\"and\'en}

\usepackage{nicefrac,comment}

\title{Subgaussian Vector Balancing by an Ising Model%
  \thanks{This research was partially supported by
    ONR Award N00014-24-12611.}}

\author{%
  Heng Guo\thanks{University of Edinburgh. Email: \texttt{hguo@inf.ed.ac.uk}.}
  \and
  Daniel A. Spielman\thanks{Yale University. Email:
    \texttt{daniel.spielman@yale.edu}}
  \and
  Xinyuan Zhang\thanks{Georgia Institute of Technology. Email:
    \texttt{normalvra@gmail.com}}%
}

\date{\today}

\begin{document}

\def\field#1{\mathbb{F}_{#1}}

\def\biE{\overrightarrow{\mathrm{E}}}

\def\Reff{\mathrm{R_{eff}}}
\def\reff{\mathrm{Reff}}
\def\Reffhat{\mathrm{\widehat{R}_{eff}}}
\def\Ceff{\mathrm{C_{eff}}}

\def\Stretch{\mathrm{Stretch}}

\def\calSO{\mathcal{SO}}

\def\cdv#1{\mathsf{cdv} (#1)}

\def\st#1#2{\mathsf{st}_{#1} (#2)}

\def\sp#1{\mathsf{sp} (#1)}
\def\Sp#1{\mathsf{sp}_{#1}}

\def\floor#1{\left\lfloor #1 \right\rfloor}
\def\ceil#1{\left\lceil #1 \right\rceil}

\def\dim#1{\mathrm{dim} (#1)}

\def\sgn#1{\mathsf{sgn} (#1)}

\def\aut#1{\mathsf{aut}(#1)}

\def\dist#1#2{\mathrm{dist} (#1, #2)}

% lect16
\def\trace#1{\mathrm{Tr} \left(#1 \right)}
\def\trop{\mathrm{Tr }}

% lect14

\def\calC{\mathcal{C}}
\def\calChat{\widehat{\mathcal{C}}}

% lect12

\def\calS{\mathcal{S}}
\def\intersect{\cap}
\def\form#1#2{\left\langle #1, #2\right\rangle}

%lect10

%\def\union{\cup}

% lect8
\def\ceil#1{\left\lceil #1 \right\rceil}
% lect 7
\def\vol#1{\mathrm{vol} (#1)}
\def\bdry#1{\partial (#1)}

% lect 6

\def\abs#1{\left|#1  \right|}

% lect 5
\def\norm#1{\left\| #1 \right\|}
\def\smallnorm#1{\| #1 \|}

\def\nucnorm#1{\left\| #1 \right\|_{*}}

\def\vs#1#2#3{#1_{#2},\ldots , #1_{#3}}

\newcommand{\energy}[1]{\mathcal{E}\left(#1 \right)}

\newcommand\zzero{\boldsymbol{0}}
\newcommand\oone{\boldsymbol{1}}

\def\calA{\mathcal{A}}
\def\calB{\mathcal{B}}
\def\calC{\mathcal{C}}
\def\calD{\mathcal{D}}
\def\calE{\mathcal{E}}
\def\calF{\mathcal{F}}
\def\calG{\mathcal{G}}
\def\calH{\mathcal{H}}
\def\calI{\mathcal{I}}
\def\calJ{\mathcal{J}}
\def\calK{\mathcal{K}}
\def\calL{\mathcal{L}}
\def\calM{\mathcal{M}}
\def\calN{\mathcal{N}}
\def\calO{\mathcal{O}}
\def\calP{\mathcal{P}}
\def\calQ{\mathcal{Q}}
\def\calR{\mathcal{R}}
\def\calS{\mathcal{S}}
\def\calT{\mathcal{T}}
\def\calU{\mathcal{U}}
\def\calV{\mathcal{V}}
\def\calW{\mathcal{W}}
\def\calX{\mathcal{X}}
\def\calY{\mathcal{Y}}
\def\calZ{\mathcal{Z}}

\renewcommand\AA{\boldsymbol{\mathit{A}}}
\newcommand\BB{\boldsymbol{\mathit{B}}}
\newcommand\CC{\boldsymbol{\mathit{C}}}
\newcommand\DD{\boldsymbol{\mathit{D}}}
\newcommand\EE{\boldsymbol{\mathit{E}}}
\newcommand\FF{\boldsymbol{\mathit{F}}}
\newcommand\GG{\boldsymbol{\mathit{G}}}
\newcommand\HH{\boldsymbol{\mathit{H}}}
\newcommand\II{\boldsymbol{\mathit{I}}}
\newcommand\JJ{\boldsymbol{\mathit{J}}}
\newcommand\KK{\boldsymbol{\mathit{K}}}
\newcommand\LL{\boldsymbol{\mathit{L}}}
\newcommand\MM{\boldsymbol{\mathit{M}}}
\newcommand\NN{\boldsymbol{\mathit{N}}}
\newcommand\OO{\boldsymbol{\mathit{O}}}
\newcommand\PP{\boldsymbol{\mathit{P}}}
\newcommand\QQ{\boldsymbol{\mathit{Q}}}
\newcommand\RR{\boldsymbol{\mathit{R}}}
\renewcommand\SS{\boldsymbol{\mathit{S}}}
\newcommand\TT{\boldsymbol{\mathit{T}}}
\newcommand\UU{\boldsymbol{\mathit{U}}}
\newcommand\VV{\boldsymbol{\mathit{V}}}
\newcommand\WW{\boldsymbol{\mathit{W}}}
\newcommand\XX{\boldsymbol{\mathit{X}}}
\newcommand\YY{\boldsymbol{\mathit{Y}}}
\newcommand\ZZ{\boldsymbol{\mathit{Z}}}

\def\aa{\pmb{\mathit{a}}}
\newcommand\bb{\boldsymbol{\mathit{b}}}
\newcommand\cc{\boldsymbol{\mathit{c}}}
\newcommand\dd{\boldsymbol{\mathit{d}}}
\newcommand\ee{\boldsymbol{\mathit{e}}}
\newcommand\ff{\boldsymbol{\mathit{f}}}
\renewcommand\gg{\boldsymbol{\mathit{g}}}
\newcommand\hh{\boldsymbol{\mathit{h}}}
\newcommand\ii{\boldsymbol{\mathit{i}}}
\newcommand\jj{\boldsymbol{\mathit{j}}}
\newcommand\kk{\boldsymbol{\mathit{k}}}
\renewcommand\ll{\boldsymbol{\mathit{l}}}

\newcommand\pp{\boldsymbol{\mathit{p}}}

\newcommand\qq{\boldsymbol{\mathit{q}}}

\newcommand\bs{\boldsymbol{\mathit{s}}}
\newcommand\nn{\boldsymbol{\mathit{n}}}
\newcommand\rr{\boldsymbol{\mathit{r}}}
\renewcommand\ss{\boldsymbol{\mathit{s}}}
\newcommand\uu{\boldsymbol{\mathit{u}}}
\newcommand\vv{\boldsymbol{\mathit{v}}}
\newcommand\ww{\boldsymbol{\mathit{w}}}
\newcommand\xx{\boldsymbol{\mathit{x}}}

\newcommand\xp{\boldsymbol{\mathit{x'}}}
\newcommand\yy{\boldsymbol{\mathit{y}}}
\newcommand\zz{\boldsymbol{\mathit{z}}}
\def\tt{\boldsymbol{\mathit{t}}}
\newcommand\mm{\boldsymbol{\mathit{m}}}

\newcommand\aatil{\boldsymbol{\widetilde{\mathit{a}}}}
\newcommand\bbtil{\boldsymbol{\widetilde{\mathit{b}}}}
\newcommand\cctil{\boldsymbol{\widetilde{\mathit{c}}}}
\newcommand\ddtil{\boldsymbol{\widetilde{\mathit{d}}}}
\newcommand\eetil{\boldsymbol{\widetilde{\mathit{e}}}}
\newcommand\fftil{\boldsymbol{\widetilde{\mathit{f}}}}
\newcommand\ggtil{\boldsymbol{\widetilde{\mathit{g}}}}
\newcommand\hhtil{\boldsymbol{\widetilde{\mathit{h}}}}
\newcommand\iitil{\boldsymbol{\widetilde{\mathit{i}}}}
\newcommand\jjtil{\boldsymbol{\widetilde{\mathit{j}}}}
\newcommand\kktil{\boldsymbol{\widetilde{\mathit{k}}}}
\newcommand\lltil{\boldsymbol{\widetilde{\mathit{l}}}}
\newcommand\mmtil{\boldsymbol{\widetilde{\mathit{m}}}}
\newcommand\nntil{\boldsymbol{\widetilde{\mathit{n}}}}
\newcommand\ootil{\boldsymbol{\widetilde{\mathit{o}}}}
\newcommand\pptil{\boldsymbol{\widetilde{\mathit{p}}}}
\newcommand\qqtil{\boldsymbol{\widetilde{\mathit{q}}}}
\newcommand\rrtil{\boldsymbol{\widetilde{\mathit{r}}}}
\newcommand\sstil{\boldsymbol{\widetilde{\mathit{s}}}}
\newcommand\tttil{\boldsymbol{\widetilde{\mathit{t}}}}
\newcommand\uutil{\boldsymbol{\widetilde{\mathit{u}}}}
\newcommand\vvtil{\boldsymbol{\widetilde{\mathit{v}}}}
\newcommand\wwtil{\boldsymbol{\widetilde{\mathit{w}}}}
\newcommand\xxtil{\boldsymbol{\widetilde{\mathit{x}}}}
\newcommand\yytil{\boldsymbol{\widetilde{\mathit{y}}}}
\newcommand\zztil{\boldsymbol{\widetilde{\mathit{z}}}}

\newcommand\aahat{\boldsymbol{\widehat{\mathit{a}}}}
\newcommand\bbhat{\boldsymbol{\widehat{\mathit{b}}}}
\newcommand\cchat{\boldsymbol{\widehat{\mathit{c}}}}
\newcommand\ddhat{\boldsymbol{\widehat{\mathit{d}}}}
\newcommand\eehat{\boldsymbol{\widehat{\mathit{e}}}}
\newcommand\ffhat{\boldsymbol{\widehat{\mathit{f}}}}
\newcommand\gghat{\boldsymbol{\widehat{\mathit{g}}}}
\newcommand\hhhat{\boldsymbol{\widehat{\mathit{h}}}}
\newcommand\iihat{\boldsymbol{\widehat{\mathit{i}}}}
\newcommand\jjhat{\boldsymbol{\widehat{\mathit{j}}}}
\newcommand\kkhat{\boldsymbol{\widehat{\mathit{k}}}}
\newcommand\llhat{\boldsymbol{\widehat{\mathit{l}}}}
\newcommand\mmhat{\boldsymbol{\widehat{\mathit{m}}}}
\newcommand\nnhat{\boldsymbol{\widehat{\mathit{n}}}}
\newcommand\oohat{\boldsymbol{\widehat{\mathit{o}}}}
\newcommand\pphat{\boldsymbol{\widehat{\mathit{p}}}}
\newcommand\qqhat{\boldsymbol{\widehat{\mathit{q}}}}
\newcommand\rrhat{\boldsymbol{\widehat{\mathit{r}}}}
\newcommand\sshat{\boldsymbol{\widehat{\mathit{s}}}}
\newcommand\tthat{\boldsymbol{\widehat{\mathit{t}}}}
\newcommand\uuhat{\boldsymbol{\widehat{\mathit{u}}}}
\newcommand\vvhat{\boldsymbol{\widehat{\mathit{v}}}}
\newcommand\wwhat{\boldsymbol{\widehat{\mathit{w}}}}
\newcommand\xxhat{\boldsymbol{\widehat{\mathit{x}}}}
\newcommand\yyhat{\boldsymbol{\widehat{\mathit{y}}}}
\newcommand\zzhat{\boldsymbol{\widehat{\mathit{z}}}}

\newcommand\AAhat{\boldsymbol{\widehat{\mathit{A}}}}
\newcommand\BBhat{\boldsymbol{\widehat{\mathit{B}}}}
\newcommand\CChat{\boldsymbol{\widehat{\mathit{C}}}}
\newcommand\DDhat{\boldsymbol{\widehat{\mathit{D}}}}
\newcommand\EEhat{\boldsymbol{\widehat{\mathit{E}}}}
\newcommand\FFhat{\boldsymbol{\widehat{\mathit{F}}}}
\newcommand\GGhat{\boldsymbol{\widehat{\mathit{G}}}}
\newcommand\HHhat{\boldsymbol{\widehat{\mathit{H}}}}
\newcommand\IIhat{\boldsymbol{\widehat{\mathit{I}}}}
\newcommand\JJhat{\boldsymbol{\widehat{\mathit{J}}}}
\newcommand\KKhat{\boldsymbol{\widehat{\mathit{K}}}}
\newcommand\LLhat{\boldsymbol{\widehat{\mathit{L}}}}
\newcommand\MMhat{\boldsymbol{\widehat{\mathit{M}}}}
\newcommand\NNhat{\boldsymbol{\widehat{\mathit{N}}}}
\newcommand\OOhat{\boldsymbol{\widehat{\mathit{O}}}}
\newcommand\PPhat{\boldsymbol{\widehat{\mathit{P}}}}
\newcommand\QQhat{\boldsymbol{\widehat{\mathit{Q}}}}
\newcommand\RRhat{\boldsymbol{\widehat{\mathit{R}}}}
\newcommand\SShat{\boldsymbol{\widehat{\mathit{S}}}}
\newcommand\TThat{\boldsymbol{\widehat{\mathit{T}}}}
\newcommand\UUhat{\boldsymbol{\widehat{\mathit{U}}}}
\newcommand\VVhat{\boldsymbol{\widehat{\mathit{V}}}}
\newcommand\WWhat{\boldsymbol{\widehat{\mathit{W}}}}
\newcommand\XXhat{\boldsymbol{\widehat{\mathit{X}}}}
\newcommand\YYhat{\boldsymbol{\widehat{\mathit{Y}}}}
\newcommand\ZZhat{\boldsymbol{\widehat{\mathit{Z}}}}

\newcommand\AAtil{\boldsymbol{\widetilde{\mathit{A}}}}
\newcommand\BBtil{\boldsymbol{\widetilde{\mathit{B}}}}
\newcommand\CCtil{\boldsymbol{\widetilde{\mathit{C}}}}
\newcommand\DDtil{\boldsymbol{\widetilde{\mathit{D}}}}
\newcommand\EEtil{\boldsymbol{\widetilde{\mathit{E}}}}
\newcommand\FFtil{\boldsymbol{\widetilde{\mathit{F}}}}
\newcommand\GGtil{\boldsymbol{\widetilde{\mathit{G}}}}
\newcommand\HHtil{\boldsymbol{\widetilde{\mathit{H}}}}
\newcommand\IItil{\boldsymbol{\widetilde{\mathit{I}}}}
\newcommand\JJtil{\boldsymbol{\widetilde{\mathit{J}}}}
\newcommand\KKtil{\boldsymbol{\widetilde{\mathit{K}}}}
\newcommand\LLtil{\boldsymbol{\widetilde{\mathit{L}}}}
\newcommand\MMtil{\boldsymbol{\widetilde{\mathit{M}}}}
\newcommand\NNtil{\boldsymbol{\widetilde{\mathit{N}}}}
\newcommand\OOtil{\boldsymbol{\widetilde{\mathit{O}}}}
\newcommand\PPtil{\boldsymbol{\widetilde{\mathit{P}}}}
\newcommand\QQtil{\boldsymbol{\widetilde{\mathit{Q}}}}
\newcommand\RRtil{\boldsymbol{\widetilde{\mathit{R}}}}
\newcommand\SStil{\boldsymbol{\widetilde{\mathit{S}}}}
\newcommand\TTtil{\boldsymbol{\widetilde{\mathit{T}}}}
\newcommand\UUtil{\boldsymbol{\widetilde{\mathit{U}}}}
\newcommand\VVtil{\boldsymbol{\widetilde{\mathit{V}}}}
\newcommand\WWtil{\boldsymbol{\widetilde{\mathit{W}}}}
\newcommand\XXtil{\boldsymbol{\widetilde{\mathit{X}}}}
\newcommand\YYtil{\boldsymbol{\widetilde{\mathit{Y}}}}
\newcommand\ZZtil{\boldsymbol{\widetilde{\mathit{Z}}}}

\newcommand\Atil{\widetilde{\mathit{A}}}
\newcommand\Btil{\widetilde{\mathit{B}}}
\newcommand\Ctil{\widetilde{\mathit{C}}}
\newcommand\Dtil{\widetilde{\mathit{D}}}
\newcommand\Etil{\widetilde{\mathit{E}}}
\newcommand\Ftil{\widetilde{\mathit{F}}}
\newcommand\Gtil{\widetilde{\mathit{G}}}
\newcommand\Htil{\widetilde{\mathit{H}}}
\newcommand\Itil{\widetilde{\mathit{I}}}
\newcommand\Jtil{\widetilde{\mathit{J}}}
\newcommand\Ktil{\widetilde{\mathit{K}}}
\newcommand\Ltil{\widetilde{\mathit{L}}}
\newcommand\Mtil{\widetilde{\mathit{M}}}
\newcommand\Ntil{\widetilde{\mathit{N}}}
\newcommand\Otil{\widetilde{\mathit{O}}}
\newcommand\Ptil{\widetilde{\mathit{P}}}
\newcommand\Qtil{\widetilde{\mathit{Q}}}
\newcommand\Rtil{\widetilde{\mathit{R}}}
\newcommand\Stil{\widetilde{\mathit{S}}}
\newcommand\Ttil{\widetilde{\mathit{T}}}
\newcommand\Util{\widetilde{\mathit{U}}}
\newcommand\Vtil{\widetilde{\mathit{V}}}
\newcommand\Wtil{\widetilde{\mathit{W}}}
\newcommand\Xtil{\widetilde{\mathit{X}}}
\newcommand\Ytil{\widetilde{\mathit{Y}}}
\newcommand\Ztil{\widetilde{\mathit{Z}}}

\newcommand\aalpha{\boldsymbol{\alpha}}
\newcommand\bbeta{\boldsymbol{\beta}}
\newcommand\ggamma{\boldsymbol{\gamma}}
\newcommand\ddelta{\boldsymbol{\delta}}
\newcommand\eepsilon{\boldsymbol{\epsilon}}
\newcommand{\zzeta}{\boldsymbol{\zeta}}
\newcommand{\eeta}{\boldsymbol{\eta}}
\newcommand{\ttheta}{\boldsymbol{\theta}}
\newcommand{\iiota}{\boldsymbol{\iota}}
\newcommand{\kkappa}{\boldsymbol{\kappa}}
\newcommand\llambda{\boldsymbol{\lambda}}
\newcommand\mmu{\boldsymbol{\mu}}
\newcommand\nnu{\boldsymbol{\nu}}
\newcommand{\xxi}{\boldsymbol{\xi}}
\newcommand\ppi{\boldsymbol{\pi}}
\newcommand{\rrho}{\boldsymbol{\rho}}
\newcommand\ssigma{\boldsymbol{\sigma}}
\newcommand\ttau{\boldsymbol{\tau}}
\newcommand\pphi{\boldsymbol{\phi}}
\newcommand{\cchi}{\boldsymbol{\chi}}
\newcommand\ppsi{\boldsymbol{\psi}}
\newcommand\oomega{\boldsymbol{\omega}}

\newcommand\GGamma{\boldsymbol{\Gamma}}
\newcommand{\DDelta}{\boldsymbol{\Delta}}
\newcommand{\TTheta}{\boldsymbol{\Theta}}
\newcommand{\LLambda}{\boldsymbol{\Lambda}}
\newcommand{\PPi}{\boldsymbol{\Pi}}
\newcommand{\SSigma}{\boldsymbol{\Sigma}}
\newcommand{\UUpsilon}{\boldsymbol{\Upsilon}}
\newcommand{\PPhi}{\boldsymbol{\Phi}}
\newcommand{\PPsi}{\boldsymbol{\Psi}}
\newcommand{\OOmega}{\boldsymbol{\Omega}}

\newcommand\Ggamma{\boldsymbol{\Gamma}}
\newcommand{\Ddelta}{\boldsymbol{\Delta}}
\newcommand{\Ttheta}{\boldsymbol{\Theta}}
\newcommand{\Llambda}{\boldsymbol{\Lambda}}
\newcommand{\Ppi}{\boldsymbol{\Pi}}
\newcommand{\Ssigma}{\boldsymbol{\Sigma}}
\newcommand{\Uupsilon}{\boldsymbol{\Upsilon}}
\newcommand{\Pphi}{\boldsymbol{\Phi}}
\newcommand{\Ppsi}{\boldsymbol{\Psi}}
\newcommand{\Oomega}{\boldsymbol{\Omega}}

\newcommand\xhat{\boldsymbol{\hat{\mathit{x}}}}
\newcommand\ahat{{\widehat{{a}}}}
\newcommand\bhat{{\widehat{{b}}}}
\newcommand\chat{{\widehat{{c}}}}
\newcommand\uhat{{\widehat{{u}}}}
\newcommand\vhat{{\widehat{{v}}}}
\newcommand\what{{\widehat{{w}}}}

\newcommand\Ghat{{\widehat{{G}}}}

\newcommand\thetahat{{\hat{\theta }}}

\newcommand\lazy{\WWtil}

% lect 4
\def\union{\cup}

% lect 3
\newdimen\pIR
\pIR= -131072sp
\newcommand\StevesR{{\rm I\kern\pIR R}}
\def\Reals#1{\StevesR^{#1}}

\def\pleq{\preccurlyeq}
\def\pgeq{\succcurlyeq}

\def\ple{\prec}
\def\pge{\succ}

% lect 2
\def\Span#1{\textbf{Span}\left(#1  \right)}
\def\bvec#1{\mathbf{#1}}

% lect 1

\def\prob#1#2{\mbox{Pr}_{#1}\left[ #2 \right]}
\def\pvec#1#2{\vec{\mbox{P}}^{#1}\left[ #2 \right]}
\def\expec#1#2{{\mathop{\mathbb{E}}}_{#1}  #2 }
\def\Expec#1#2{{\mathop{\mathbb{E}}}_{#1}\left( #2 \right)}
\def\var#1{\mbox{\bf Var}\left( #1 \right)}
\def\varh#1#2{\mbox{\bf Var}_{#1}\left( #2 \right)}
\def\covh#1#2{\mbox{\bf Cov}_{#1}\left( #2 \right)}

\def\symExpec{\mathbb{E}}

\def\defeq{\stackrel{\mathrm{def}}{=}}
\def\setof#1{\left\{#1  \right\}}
\def\mand{\mbox{ and }}
\def\mor{\mbox{ or }}

\def\sizeof#1{\left|#1  \right|}

\newsavebox{\Codefig}

\newenvironment{codefig}[1]
{
\sbox{\Codefig}{\centering\epsfig{figure=figs/#1,width=4in}}%
\noindent
\begin{minipage}{.3\linewidth}
\begin{alltt}}
{\end{alltt}
\end{minipage}\hfill
\begin{minipage}{.6\linewidth}
\usebox{\Codefig}
\end{minipage}
}

\newenvironment{codeimg}[1]
{
\sbox{\Codefig}{\centering\epsfig{figure=figs/#1,width=4in}}%
\begin{center}
\begin{minipage}{.7\linewidth}
\usebox{\Codefig}
\begin{alltt}}
{\end{alltt}
\end{minipage}
\end{center}
}

\newenvironment{codeimg_label}[2]
{
\sbox{\Codefig}{\centering\epsfig{figure=figs/#1,width=4in}}%
\begin{center}
\begin{minipage}{.7\linewidth}
\usebox{\Codefig}
\begin{alltt}}
{\end{alltt}
\end{minipage}
\end{center}
}

\newenvironment{codefigRev}[1]
{
\sbox{\Codefig}{\centering\epsfig{figure=figs/#1,width=4in}}%
\noindent
\begin{minipage}{.3\linewidth}
\begin{alltt}}
{\end{alltt}
\end{minipage}\hfill
\begin{minipage}{.6\linewidth}
\usebox{\Codefig}
\end{minipage}
}

\newenvironment{codefig2}[1]
{
\sbox{\Codefig}{\centering\epsfig{figure=figs/#1,height=2.5in}}%
\noindent
\begin{minipage}{.45\linewidth}
\begin{alltt}}
{\end{alltt}
\end{minipage}\hfill
\begin{minipage}{.45\linewidth}
\usebox{\Codefig}
\end{minipage}
}

\newenvironment{codefig3}[1]
{
\sbox{\Codefig}{\centering\epsfig{figure=figs/#1,width=.45\linewidth}}%
\noindent
\begin{minipage}{.45\linewidth}
\begin{alltt}}
{\end{alltt}
\end{minipage}\hfill
\begin{minipage}{.45\linewidth}
\usebox{\Codefig}
\end{minipage}
}

\def\vecE{\vec{E}}

\def\id{\mathit{id}}

\def\lprod{{\ooalign{\hfil\raise .00ex\hbox{\scriptsize L}\hfil\crcr\mathhexbox20D}}}

\def\acos{\operatorname{acos}}
\def\acosh{\operatorname{acosh}}
\def\asinh{\operatorname{asinh}}

\def\csum#1#2{#1 \{ #2 \}}  
  
\def\match#1{\mu_{x}\left[ #1 \right]}

\newcommand{\verdiere}{Verdi{\`e}re }
\newcommand{\lovasz}{Lov{\`a}sz }
\newcommand{\poincare}{Poincar{\'e}}
\newcommand{\erdos}{Erd{\"o}s}
\newcommand{\renyi}{R{\'e}nyi}

\newcommand{\lecture}[3]{
    %!LW recipe=pdflatex
\chapter{#3}
    \begin{center}
        Lecture #1 from #2
    \end{center}
}

\maketitle

\begin{abstract}
We show that the guaranties of the Gram-Schmidt Walk Algorithm of Bansal, Dadush, Garg, and Lovett can be achieved by an Ising model.
Given vectors $\vv_1, \ldots, \vv_n$ of norm at most 1, we give a closed form for a centrally symmetric distribution on a vector 
$\ssigma \in \setof{\pm 1}^n$ so that the signed sum 
$\sum_j \ssigma(j) \vv_j$ is $1$-subgaussian.
We show that the Glauber dynamics for this model mix rapidly provided that the norms of the input vectors are bounded below 1 by an inverse-polynomial.

The proofs employ the theory of Lorentzian polynomials, and we use these techniques to show fast mixing bounds for other spin systems.
\end{abstract}

\thispagestyle{empty}
\clearpage
\pagenumbering{arabic}

\newpage

\section{Introduction}

Dadush, Garg, Lovett, and Nikolov~\cite{dadush2019towards} proved that Banaszczyk’s Vector Balancing Theorem~\cite{banaszczyk1998balancing} is equivalent to the following sampling problem:
given vectors $\vv_1, \ldots, \vv_n$ in the unit ball, sample a vector 
$\ssigma \in \setof{\pm 1}^n$ so that the signed sum 
$\sum_j \ssigma(j) \vv_j$ has expectation zero and is $C$-subgaussian, for some constant $C$.
We call an algorithm that achieves this guarantee a \textit{$C$-subgaussian vector balancing algorithm}.
Bansal, Dadush, Garg and Lovett~\cite{BansalDGL19} developed a polynomial-time algorithm, called the Gram-Schmidt Walk, that solves this sampling problem and thereby provides a polynomial-time realization of Banaszczyk's Theorem.
Harshaw, S{\"a}vje, Spielman and Zhang~\cite{harshaw2024balancing} proved that the constant achieved by this algorithm is actually $1$, which is the best possible. 

We show that one can obtain a 1-subgaussian vector balancing algorithm from an Ising model that encodes a very simple distribution, and that the natural Glauber dynamics sample from this model in polynomial-time.
If all the vectors have Euclidean norm at most 1, then the distribution samples $\ssigma$ with probability proportional to 

\begin{align}\label{eqn:weight}
  w(\ssigma) & \defeq\prod_{1\le i<j\le n}\left(1-\ssigma(i) \ssigma(j) \inner{}{\vv_i,\vv_j}\right).
\end{align}
This is an Ising model with interaction matrix $\JJ$ given by $\JJ(i,j) = \frac{1}{2} \log \frac{1+\inner{}{\vv_i,\vv_j}}{1-\inner{}{\vv_i,\vv_j}} = \mathrm{atanh}(\inner{}{\vv_i,\vv_j})$ because $w(\ssigma)$ is proportional to $\exp(-\ssigma^T \JJ \ssigma/ 2)$.

This distribution is well-defined so long as the input does not contain any three parallel unit vectors.
If multiple input vectors have the form $\pm \vv$ where $\vv$ has norm 1, then we fix their sum to be zero if there are an even number of them, or to $\pm \vv$ if their number is odd.
The mixing time of the Glauber dynamics depends on the reciprocal of 1 minus the maximum norm of a vector.
To efficiently sample when some vectors have norm 1 or very close to it, we scale down the vectors and use rejection sampling.  See Section~\ref{sec:boundary_cases} for the details.

Our motivation for this paper is three-fold.
First, we believe that 1-subgaussian vector balancing is sufficiently fundamental that we should know if it can be done with a simple and cleanly described distribution. 
To the best of our knowledge, ours is the first closed form $C$-subgaussian sampler for any $C$.
Second, if one wants to vary the vector balancing guarantee, it should be much easier to modify the distribution directly than it would be to modify the Gram-Schmidt Walk.
Third, we expect that experimenters will be more likely to use 1-subgaussian samplers to design randomized controlled trials, as suggested in~\cite{harshaw2024balancing}, if they have a better understanding of the distribution.

We now state our main results.

\begin{theorem}\label{thm:subgaussian}
Assume that the vectors $\vv_1, \ldots, \vv_n$ all have Euclidean norm at most 1
  and that no three are parallel unit vectors.
Let $\ssigma$ be chosen with probability proportional to 
  $w(\ssigma)$.
  Let $\ss = \sum_i \ssigma(i) \vv_i$.
  Then, $\ss$ is a 1-subgaussian random vector.
  That is, for all vectors $\ttheta$, 
  \[
  \expec{}{\exp (\inner{}{\ttheta, \ss})}
  \leq \exp(\norm{\ttheta}^2 / 2).  
  \]
\end{theorem}

Theorem~\ref{thm:subgaussian} is proved as a special case of Theorem~\ref{thm:subgaussianh}, which shows that the same holds under arbitrary external fields after re-centering $\ss$.

We prove rapid mixing of the Glauber dynamics for the Ising distribution~\eqref{eqn:weight},
when the norms of the input vectors are inverse-polynomially bounded away from $1$. 
This also holds under external fields.
% \begin{theorem}\label{thm:spectral-gap}
%   For every $h\in\=R^n$, let $P_h$ be the Glauber dynamics for $\mu_h$, where $\mu_h(\ssigma) \propto \exp\tp{\inner{}{h,\ssigma}}w(\sigma)$ denotes the Ising model under external fields $h$.
%   Let $c_0\defeq 1-\max_{i\in[n]}\norm{\vv_i}^2$. The spectral gap of $P_h$ satisfies
%   \begin{align*}
%     \operatorname{gap}(P_h)\ge\frac{c_0}{n}.
%   \end{align*}
%   Consequently, for every $0<\varepsilon<1$, its mixing time satisfies
%   \begin{align*}
%     t_{\mathrm{mix},h}(\varepsilon)
%     =O\left(\frac{n}{c_0}\left(
%       \log\frac1\varepsilon+\norm{h}_1
%       +n^2\log c_0^{-1}
%     \right)\right).
%   \end{align*}
% \end{theorem}

\begin{theorem}\label{thm:spectral-gap}
  % For every $h\in\=R^n$, let $P_h$ be the Glauber dynamics for $\mu_h$, where $\mu_h(\ssigma) \propto \exp\tp{\inner{}{h,\ssigma}}w(\sigma)$ denotes the Ising model under external fields $h$.
  Let $\alpha_0\defeq 1-\max_{i\in[n]}\norm{\vv_i}^2$. The spectral gap of the Glauber dynamics $P$ for the Ising distribution~\eqref{eqn:weight} satisfies
  \begin{align*}
    \operatorname{gap}(P)\ge\frac{\alpha_0}{n}.
  \end{align*}
  Consequently, for every $0<\varepsilon<1$, its mixing time satisfies
  \begin{align*}
    t_{\mathrm{mix}}(\varepsilon)=O\left(\frac{n}{\alpha_0}\left(\log\frac1\varepsilon+n^2\log \alpha_0^{-1}\right)\right).
  \end{align*}
\end{theorem}

This is not quite as fast as
  the Gram-Schmidt Walk~\cite{BansalDGL19}, which can be implemented to run in time $O(n^2 d)$.
We conjecture that our distribution satisfies a modified log-Sobolev inequality with constant $O(\alpha_0 / n)$, which would get us close to this running time bound: it would imply mixing in time $O((n / \alpha_0) \log (n \log(1 / \alpha_0) / \epsilon))$.
Those who are willing to relax the subgaussian bound can obtain a $\sqrt{3}$-subgaussian guarantee in time 
  $O(n d)$, although again without a closed form distribution~\cite{adenali2026optimalonlinediscrepancyminimization}.

To sample from the distribution~\eqref{eqn:weight}, there is a faster way than using the mixing time guarantee, namely the checkpoint technique in \cite[Section 5]{CGVY26v1}.
Introduce a temperature parameter, and (non-adaptively and deterministically) find an $\widetilde{O}(\sqrt{n})$-length cooling schedule with bounded $\chi^2$-divergence between two adjacent temperatures.
Then, run the Glauber dynamics at each temperature, and once it is mixed, use the last state as a warm start for the next temperature.
Overall we only need $\widetilde{O}(n^{1.5}/\alpha_0)$ Markov chain steps.
Since this is a straightforward and routine application of  \cite[Section 5]{CGVY26v1}, we refer interested readers to read that and omit the details here.

In Section~\ref{sec:boundary_cases} we give a few ways of modifying the distribution and/or the sampling procedure to handle vectors of norm up to 1. 

\Cref{thm:spectral-gap} provides a new spectral criterion for the rapid mixing of Glauber dynamics on general Ising models.
Consider the Ising distribution on $\{\pm 1\}^n$ specified by a symmetric interaction matrix $\JJ \in \Reals{n \times n}$ with zero diagonal and an arbitrary external field $\hh \in \Reals{n}$:
\begin{align}\label{eqn:general-Ising}
  \forall \ssigma \in \{\pm 1\}^n, \qquad \mu(\ssigma) \propto \exp\tp{\frac{1}{2} \transpose{\ssigma} \JJ \ssigma + \transpose{\hh} \ssigma}.
\end{align}

\begin{theorem}\label{thm:spectral-gap-general-ising}
  Let $\JJ \in \Reals{n \times n}$ be a symmetric matrix with zero diagonal, let $\hh \in \Reals{n}$ be an arbitrary external field,
  and let $\TT = \tanh(\JJ) \defeq (\tanh(\JJ(i,j)))_{i,j\in[n]}$ denote the entrywise hyperbolic-tangent matrix of $\JJ$.
  If $\lambda_{\max}(\TT) < 1$, then the spectral gap of the Glauber dynamics $P$ for the Ising distribution~\eqref{eqn:general-Ising} satisfies
  \begin{align*}
    \operatorname{gap}(P) \ge \frac{1-\lambda_{\max}(\TT)}{n}.
  \end{align*}
\end{theorem}

\begin{remark}
For general Ising models, previous work~\cite{EKZ22,AJKPV22} established the rapid mixing of Glauber dynamics when the interaction matrix satisfies $0 \preceq \JJ \preceq (1-\epsilon) \II$ for some $\epsilon > 0$.
The conditions of \Cref{thm:spectral-gap-general-ising} cover this as a special case.
To see this, let $\TT$ have zero diagonal and off-diagonal entries $\TT(i,j)=\tanh \JJ(i,j)$.
Since diagonal entries do not affect the Ising distribution, it suffices by \Cref{thm:spectral-gap-general-ising} to show that $\HH \defeq \JJ - \TT \succeq 0$.
This follows from the fact that $\HH$ is symmetric, has non-negative diagonals, and is diagonally-dominant: for all $i \in [n]$, 
\begin{align*}
  \sum_{j \neq i} \abs{\HH(i,j)} = \sum_{j \neq i} \abs{\JJ(i,j) - \tanh \JJ(i,j)} < \sum_{j \neq i} \abs{\JJ(i,j)}^3 \le \sum_{j \neq i} \abs{\JJ(i,j)}^2 \le (\JJ^2)_{i,i} \le \JJ(i,i) = \HH(i,i),
\end{align*}
where we use $\abs{x - \tanh x} \le \abs{x}^3 / 3$, $\abs{\JJ(i,j)}\le \norm{\JJ}_{\mathrm{op}}< 1$, $\JJ \succeq 0$ and $\JJ^2 \preceq \JJ$.
\end{remark}

We apply \Cref{thm:spectral-gap-general-ising} to the Sherrington--Kirkpatrick model, which corresponds to the Ising distribution~\eqref{eqn:general-Ising} with symmetric interaction matrix $\JJ \in \Reals{n \times n}$ given by independent entries $\JJ(i,j) \sim \+N(0,\beta^2/n)$.
The high-probability bound $\lambda_{\max}(\TT) = 2\beta + o_n(1)$ yields the following guarantee.

\begin{corollary}\label{cor:SK-spectral-gap}
  For any $0 \le \beta < 1/2$ and any external field $\hh \in \Reals{n}$, with probability $1-o_n(1)$ over the Gaussian couplings $\JJ$, the Glauber dynamics $P$ for the Sherrington--Kirkpatrick model satisfies
  \begin{align*}
    \operatorname{gap}(P) \ge \frac{1 - 2\beta - o_n(1)}{n}.
  \end{align*}
\end{corollary}

The proofs of~\Cref{thm:spectral-gap-general-ising} and~\Cref{cor:SK-spectral-gap} are given at the end of~\Cref{sec:Glauber}.
Previously, the spectral radius criterion~\cite{EKZ22,AJKPV22} established a spectral gap of order $\Omega(1/n)$ for $\beta < 1/4$, which was subsequently improved to $\beta \approx 0.295$ using the trickle-down theorem~\cite{AKV24}.
Recently, remarkable progress has been achieved in sampling the Sherrington--Kirkpatrick model, including total-variation sampling and warm-start mixing of Glauber dynamics for the zero-field model at $\beta < 1/2$~\cite{DLSS26,DLSS26a} as well as stronger worst-start mixing guarantees~\cite{Wang26,BLO26}.
These results were further extended across the entire high-temperature regime $\beta < 1$~\cite{LSS26,HLY26,DHMR26,OAI26}.

This approach also yields spectral gap bounds for general spin systems.
Specifically, we establish rapid mixing of Glauber dynamics for anti-ferromagnetic spin systems on complete graphs whose interaction matrices have strictly positive diagonal entries, and derive spectral gap bounds in terms of the interaction matrix and eigenvalues of the graph's adjacency matrix.
For the anti-ferromagnetic Potts model, our results extend the parameter range and improve the spectral gap lower bound obtained in recent work of Chen and Wang~\cite{CW26}. These extensions are presented in~\Cref{sec:general}.

\section{Preliminaries}

% Recall subgaussian, and other defs.

% Define $[n]$.

Let $[n] = \{1,2,\ldots,n\}$. A random vector $X \in \mathbb{R}^d$ is $C$-subgaussian if for all vectors $\ttheta \in \mathbb{R}^d$,
\begin{align*}
  \expec{}{\exp \left(\inner{}{\ttheta, X - \expec{}{X}}\right)} \le \exp\left(C^2 \norm{\ttheta}^2/2\right).
\end{align*}

Let $P$ be a reversible Markov chain on a finite state space $\Omega$ with stationary distribution $\mu$, that is, $\mu(\ssigma) P(\ssigma,\ttau) = \mu(\ttau) P(\ttau,\ssigma)$ for all $\ssigma,\ttau \in \Omega$.
The spectral gap of the Markov chain $P$ is defined as the difference between $1$ and the second largest eigenvalue of $P$, and for $0 < \varepsilon < 1$, the mixing time $\mixingtime{\varepsilon}$ is given by
\begin{align*}
  \mixingtime{\epsilon} \defeq \max_{\ssigma \in \Omega} \min \left\{t \in \mathbb{N} \mid \DTV{P^t(\ssigma,\cdot)}{\mu} < \epsilon\right\},
\end{align*}
where $\DTV{\mu}{\nu}$ denotes the total variation distance between distributions $\mu$ and $\nu$.

Let $\mu$ be a distribution over a finite product space $\mathcal{Q} = \prod_{i=1}^n Q_i$.
The Glauber dynamics for $\mu$ updates a feasible configuration $\ssigma \in \mathcal{Q}$ by selecting a coordinate $i \in [n]$ uniformly at random, and replaces $\ssigma_i$ by $c \in Q_i$ with probability proportional to $\mu(\ssigma^{i \gets c})$.
Here, $\ssigma^{i \gets c}$ is the configuration obtained from $\ssigma$ by replacing its $i$-th coordinate with $c$ and leaving all other coordinates untouched.
The Glauber dynamics is reversible with respect to $\mu$.

\section{Motivating the distribution}

We came to the distribution \eqref{eqn:weight} by a semi-rigorous process.
We enumerated a few properties that we thought a distribution solving this problem should have, enumerated a couple of distributions that had those properties, and then eliminated those that failed to produce 1-subgaussian distributions in small experiments.
This was done with conventional computation before the release of useful LLMs.

If we only have two input vectors and both of them are unit vectors, then the distribution \eqref{eqn:weight} minimizes the operator norm of the covariance matrix of the resulting distribution.
As this seems desirable, we sought distributions that agreed with this one when given only two vectors.

When the input is comprised of many identical small vectors in one dimension, say $\vv_i = \epsilon$ for all $i$,  the induced distribution resembles a discrete Gaussian: as $n$ grows, the probability of obtaining the sum $k \epsilon$ becomes approximately proportional to $\exp(-(k \epsilon)^2 / 2)$, for $k$ equivalent to $n$ modulo $2$.

Finally, we wanted to be sure that entirely orthogonal sets of vectors would not interact, and that orthogonal pairs of vectors would not interact directly.

\section{Handling vectors of norm up to 1}\label{sec:boundary_cases}

The distribution given by \eqref{eqn:weight} is well-defined so long as there is a $\ssigma$ for which $w(\ssigma)$ is non-zero.
And, the only way that $w(\ssigma)$ can be uniformly zero is if the input contains at least three parallel unit vectors.
We can remove such cases from consideration by deterministically setting the signs of some of those vectors.
Given an even number of parallel unit vectors, we can set their signs so that their sum is zero, and thus does not contribute to the signed sum.
Given an odd number that is larger than three, we can remove all but one by fixing the signs of all but one so that they sum to zero.

While this transformation ensures that the distribution is well-defined and that the 1-subgaussian bound holds, it does not resolve the problem that our bound on the mixing time of Glauber dynamics depends on the reciprocal of 1 minus the maximum norm of an input vector.
We suggest two ways of addressing this problem.

A crude solution is to
  divide all the input vectors by $1 + \epsilon$ for some small $\epsilon$.
The Glauber dynamics will mix in time depending on $1/\epsilon$, and the resulting distribution on the \textit{original} vectors is guaranteed to be at most $(1+\epsilon)$-subgaussian.
This will be enough for most practical purposes.

A principled way to address the problem while preserving the 1-subgaussian guarantee begins by 
observing that division of vectors by $(1+\epsilon)$ does not substantially change the distribution if no input vectors have inner product close to $1$ or $-1$.
And, we can remove pairs of vectors with such inner products by replacing them with their sum or difference.

Let $c = \sqrt{3} - 1.$
If any pair of input vectors $\vv_i$ and $\vv_j$ have
  $\inner{}{\vv_i, \vv_j} > c $, then $\vv_i - \vv_j$ has norm less than $c$.
The same holds for $\vv_i + \vv_j$ if $\inner{}{\vv_i, \vv_j} < -c $.
After performing these replacements, we will be left with an input in which 
  all pairs of vectors have inner products of absolute value at most $c$.
After sampling from the distribution on this smaller input, we can replace the new vectors with with the sum or difference that produced them.
This will ensure that the 1-subgaussian guarantee is satisfied.

To sample from the resulting distribution, we can now sample from the distribution obtained by dividing all the vectors by $(1 + \epsilon)$, where $\epsilon = 1/(4 n^2)$, and using rejection sampling
  (see \cite[Chapter 2, Section 3]{Devroye1986}).
Let $w_{\epsilon}(\ssigma)$ the distribution obtained after this division.
One can show that for every $\ssigma$, 
\[
1/2 w_{\epsilon}(\ssigma) \leq  w(\ssigma) \leq 2 w_{\epsilon}(\ssigma).
\]
After sampling from the modified distribution, we return the sampled  $\ssigma$ with probability $w(\ssigma) / 2 w_{\epsilon}(\ssigma)$.
Otherwise, we repeat this procedure until we do return a $\ssigma$.
As the probability of success is at least $1/4$, we expect to need at most 4 samples from the distribution proportional to $w_{\epsilon}$.

\section{A discrete Alexandrov-Fenchel inequality and Lorentzian Polynomials}

The key to our proof of Theorem~\ref{thm:subgaussian} is a discrete analog of the 
  Alexandrov-Fenchel inequality \cite{Fen36,Ale37} 
  proved by 
  Lee, Oh, and Seo \cite{LOS25} using the theory of Lorentzian Polynomials~\cite{BH20}.
Lorentzian polynomials, under the name \textit{strongly log-concave polynomials}~\cite{Gurvits09} have been used to analyze mixing rates of Markov chains \cite{CGM21, ALOV24}.
However, their use in our paper through Lee, Oh and Seo's discrete Alexandrov-Fenchel inequality appears novel.

We will say that a symmetric matrix is \textit{Lorentzian} if its entries are non-negative and it has at most one positive eigenvalue\footnote{
If a non-negative symmetric matrix has a positive entry, then it has at least one positive eigenvalue. So, the only Lorentzian matrices with no positive eigenvalues are the all-zero matrices.}.
Given a finite set of coordinates, $C$, 
a Lorentzian matrix $\AA \in \Reals{C \times C}$, 
and $k$ vectors 
 $\xx_1, \ldots, \xx_k \in \Reals{C}$, Lee, Oh and Seo examine the following polynomial whose variables are the entries of those vectors:

\begin{equation}\label{eqn:defV}
  \calV_{\AA}(\xx_1,\dots,\xx_k) \defeq 
  \sum_{c:[k]\rightarrow C} \prod_{i=1}^k \xx_i(c(i))\prod_{\setof{i,j} \in \binom{[k]}{2}} \AA(c(i),c(j)).
\end{equation}

The following theorem may be derived from Theorem~3.1, Proposition~4.1, and the discussion following it through equation (4.1) of their paper~\cite{LOS25}:
\begin{theorem} \label{thm:AF}
  Let $\AA \in \Reals{C \times C}$ be a Lorentzian matrix.
  Then, for all vectors $\xx_1 \in \Reals{C}$ and 
  $\xx_2,\dots,\xx_{k}\in \Reals{C}_{\ge 0}$,
  we have 
  \begin{align}    \label{eqn:AF}
    \+V_{\AA}(\xx_1,\xx_2,\xx_3,\dots,\xx_k)^2 \ge 
    \+V_{\AA}(\xx_1,\xx_1,\xx_3,\dots,\xx_k) 
    \+V_{\AA}(\xx_2,\xx_2,\xx_3,\dots,\xx_k).
  \end{align}
\end{theorem}

To prove this, Lee, Oh, and Seo show that $ \+V_{\AA}(\xx_1,\dots,\xx_n)$ is the polarization of a Lorentzian polynomial and then apply results of \branden\ and Huh.
They point out that their Proposition 4.1 is in fact Proposition 4.5 of \cite{BH20}.
The first inequality of this type of which we are aware appears in the work of Van Lint \cite[Lemma 2.5]{van1981notes}.

Two types of vectors will arise in our proofs: the vectors $\vv_1, \ldots \vv_n$ that we want to balance and the vectors 
$\xx_1,\dots,\xx_k$ that appear in the definition of $\calV_{\AA}$.
To avoid confusion, 
we will call the first set the \textit{input vectors}
and the second set the \textit{selection vectors}.

We define the \textit{Lorentz matrix} of vectors $\setof{\vv_c}_{c \in C}$ to be the matrix $\AA$ with $\AA (c, \chat) =1 - \inner{}{\vv_c, \vv_{\chat}} $.
The following lemma is the first step to using Theorem~\ref{thm:AF}
  to analyze the distribution \eqref{eqn:weight}.

\begin{lemma}\label{lem:vec_to_mat}
Let $\AA$ be the Lorentz matrix of vectors $\setof{\vv_c}_{c \in C}$.
If all $\vv_c$ have norm at most one, 
then $\AA$ is a Lorentzian matrix.    
\end{lemma}

\begin{proof}
Let $\JJ \in \Reals{C \times C}$ be the all-1 matrix and let 
$\VV \in \Reals{d \times C}$ be the matrix whose columns are the vectors $\vv_c$. 
Then
\[
  \AA = \JJ - \VV^T \VV .
\]
As all the inner products of the vectors are at most $1$, the matrix $\AA$ is non-negative. 
As $\VV^T \VV$ is positive semidefinite,
  $\zz^T \AA \zz \leq 0$ for all $\zz$ orthogonal to $\oone$.
This implies that $\AA$ has at least $\sizeof{C}-1$ non-positive eigenvalues, and so it has at most one positive eigenvalue.
\end{proof}

\section{Subgaussian concentration}

To relate the distribution of $\ss$ to Theorem \ref{thm:AF}, we consider the Lorentz matrix of the input vectors and their opposites.
To index these, let $\calC = \setof{1^+, 1^-, 2^+, 2^-, \ldots, n^+, n^-}$, and set $\vv_{i^+} = \vv_i$ and $\vv_{i^-} = -\vv_i$.
Let $k = n$ and for $1 \leq i \leq n$ define the selection vector $\xx_i$ so that
\[
\xx_i (c) = \begin{cases}
    1 & \text{if $c = i^+$}\\
    1 & \text{if $c = i^-$}\\
    0 & \text{otherwise.}
\end{cases}
\]

With these selection vectors, all nonzero terms in \eqref{eqn:defV}
  must have $c(i) \in \setof{i^+, i^-}$ for all $i$.
Thus, the $c : [n] \rightarrow \calC$ that can lead to nonzero terms are in bijection with 
  $\ssigma \in \setof{\pm 1}^n$, and
\begin{equation*}
\calV_{\AA} (\xx_1, \ldots, \xx_n) = 
  \sum_{\ssigma \in \setof{\pm 1}^n} w(\sigma) 
  \defeq \calZ,
\end{equation*}
where $\calZ$ is called the \textit{partition function} of the distribution.

We will need to consider generalizations of the distribution given by $w$ that have been ``tilted'' by an ``external field'' $\hh \in \Reals{n}$.
We define 
\begin{equation}\label{eqn:tilted-weight}
 w_{\hh} (\ssigma) = e^{\inner{}{\hh, \ssigma}} w (\ssigma),
\end{equation}
and 
$\calZ(\hh) = \sum_{\ssigma \in \setof{\pm 1}^n} w_{\hh}(\sigma) $.
Thus, we sample $\ssigma$ with probability 
\begin{equation}\label{eqn:mu}
 \mu_{\hh} (\ssigma) = \frac{ w_{\hh}(\ssigma) }{\calZ(\hh)}.
\end{equation}
We write $\expec{\hh}{}$ and $\mathbf{Var}_{\hh}$ to indicate expectation and variance when $\ssigma$ is distributed according to this distribution.
The assumption that there are no triple of parallel unit vectors ensures that $\calZ(\hh)$ does not equal zero and thus these distributions are well-defined.

We can express $\mu_{\hh} (\ssigma)$ in terms of $\calV_{\AA}$ by changing the selection vectors.
Upon changing $\xx_i$ so that 
\begin{equation}\label{eqn:h_selection}
\xx_i (c) = \begin{cases}
    e^{\hh(i)} & \text{if $c = i^+$}\\
    e^{-\hh(i)} & \text{if $c = i^-$}\\
    0 & \text{otherwise,}
\end{cases}
\end{equation}
we obtain 
\[
\calV_{\AA} (\xx_1, \ldots, \xx_n) = \calZ(\hh).
\]

We now upper bound the covariance matrix of the signed sum 
  $\ss$ for every external field. 

\begin{lemma} \label{lem:cov}
Assume that the vectors $\vv_1, \ldots, \vv_n$ all have Euclidean at most 1 and that no three are parallel unit vectors.
Let $h \in \Reals{n}$, 
  $\ssigma$ be distributed according to $w_{\hh}(\ssigma) / \calZ(\hh)$, 
  and $\ss = \sum_i \ssigma(i) \vv_i$.
Then for every $\ttheta \in \Reals{d}$,
\[
 \var{\inner{}{\ttheta, \ss}} \leq \norm{\ttheta}^2.
\]    
\end{lemma}

We begin by introducing the notation we use in the proof.
We will consider the functions $\calV_{\BB}$ where $\BB$ is the Lorentz matrix of the input vectors, their negatives,
  the all-zero vector $\zzero$,
  and a vector $\epsilon \ttheta$.
We index vector $\zzero$ by coordinate 0 and vector $\epsilon \ttheta$ by coordinate $-1$.
We will set $k = n+2$, and consider two more selection vectors $\xx_0$ and $\xx_{-1}$ that we define to be the elementary unit vectors in directions $0$ and $-1$.
Concretely, we are instantiating \eqref{eqn:defV} by setting $\calChat = \setof{-1,0} \union \calC$, $K = \setof{-1, 0, 1, \ldots, n}$ and taking
\[
  \calV_{\BB}(\yy_{-1}, \yy_{0}, \yy_1 ,\dots,\yy_n) \defeq 
  \sum_{c:K \rightarrow \calChat} \prod_{i \in K} \xx_i(c(i))\prod_{\setof{i,j} \in \binom{K}{2}} \BB(c(i),c(j)).
  \]

For selection vectors $\aa$ and $\bb$, each of which will be one of $\xx_0$ or $\xx_{-1}$, we define
\[
\calZ (\aa, \bb) \defeq \calV_{\BB} (\aa, \bb, \xx_1, \ldots, \xx_n),
\]
where we keep $\xx_1, \ldots, \xx_n$ as they were defined in \eqref{eqn:h_selection}.

\begin{proof}[Proof of Lemma~\ref{lem:cov}]
Let $\epsilon > 0$ satisfy $\epsilon \norm{\ttheta} < 1$.
We will eventually let $\epsilon$ go to zero.

Theorem~\ref{thm:AF} tells us that 
\begin{equation}\label{eqn:useAF}
     \calZ (\xx_0, \xx_{-1})^2 \geq 
 \calZ (\xx_0, \xx_0) \calZ(\xx_{-1}, \xx_{-1}).
\end{equation}

We now expand the terms that appear in this inequality.
As $\xx_0$ pins a coordinate to $0$ and that coordinate indexes the all-zero vector, 
\[
  \calZ (\xx_0, \xx_0)  = 
   \sum_{\ssigma \in \setof{\pm 1}^n} w_{\hh}(\ssigma)
 (1 - \inner{}{\zzero, \zzero}) 
 \prod_{i = 1}^n (1 - \inner{}{\zzero, \ssigma(i) \vv_i}) 
 \prod_{i = 1}^n (1 - \inner{}{\zzero, \ssigma(i) \vv_i}) 
  = \calZ(\hh).
\]
Similarly 
\begin{align*}
  \calZ (\xx_0, \xx_{-1})
   & =  
      \sum_{\ssigma \in \setof{\pm 1}^n} w_{\hh}(\ssigma)
 (1 - \inner{}{\zzero, \epsilon \ttheta}) 
 \prod_{i = 1}^n (1 - \inner{}{\zzero, \ssigma(i) \vv_i}) 
 \prod_{i = 1}^n (1 - \inner{}{\epsilon \ttheta, \ssigma(i) \vv_i}) 
 \\ & = 
    \sum_{\ssigma \in \setof{\pm 1}^n} w_{\hh}(\ssigma)
   \prod_{i = 1}^n (1 - \epsilon \inner{}{\ttheta, \ssigma(i) \vv_i}) ,
   \qquad \text{and}
   \\
  \calZ (\xx_{-1}, \xx_{-1})
   & =  
      \sum_{\ssigma \in \setof{\pm 1}^n} w_{\hh}(\ssigma)
 (1 - \inner{}{\epsilon \ttheta, \epsilon \ttheta}) 
 \prod_{i = 1}^n (1 - \inner{}{\epsilon \ttheta, \ssigma(i) \vv_i}) 
 \prod_{i = 1}^n (1 - \inner{}{\epsilon \ttheta, \ssigma(i) \vv_i}) 
 \\ & = 
          (1 - \epsilon^2 \inner{}{\ttheta, \ttheta})
    \sum_{\ssigma \in \setof{\pm 1}^n} w_{\hh}(\ssigma)
   \left( 
   \prod_{i = 1}^n (1 - \epsilon \inner{}{\ttheta, \ssigma(i) \vv_i})
   \right)^2 .      
\end{align*}

Define 
\[
\PPhi (\ssigma) =    
  \prod_{i = 1}^n (1 - \epsilon \inner{}{\ttheta, \ssigma(i) \vv_i}) ,
\]
so when we choose $\ssigma$ with probability
  $w_{\hh}(\ssigma) / \calZ(\hh)$,
  we can write
\begin{align*}
      \calZ (\xx_{0}, \xx_{-1})
    & = \calZ(\hh) \expec{\hh}{ \PPhi(\ssigma)  }, \qquad \text{and}
    \\
 \calZ (\xx_{-1}, \xx_{-1})
    & = (1 - \epsilon^2 \inner{}{\ttheta, \ttheta}) 
     \calZ(\hh) \expec{\hh}{ \PPhi(\ssigma)^2  }.
\end{align*}
Substituting these identities into \eqref{eqn:useAF}
and  canceling $\calZ(\hh)^2$ from both sides yields
\begin{align*}
\left( \expec{\hh}{ \PPhi(\ssigma)  }  \right)^2 
& \geq (1 - \epsilon^2 \inner{}{\ttheta, \ttheta}) 
      \expec{\hh}{ \PPhi(\ssigma)^2  } \qquad \iff 
      \\
\epsilon^2 \norm{\ttheta}^2 
      \expec{\hh}{ \PPhi(\ssigma)^2  }
& \geq 
  \expec{\hh}{ \PPhi(\ssigma)^2  } - \left( \expec{\hh}{ \PPhi(\ssigma)  }  \right)^2 
\\
&  = \varh{\hh}{\PPhi(\ssigma)},
\end{align*}
which implies
\[
\norm{\ttheta}^2 
      \expec{\hh}{ \PPhi(\ssigma)^2  }
      \geq 
      \frac{\varh{\hh}{\PPhi(\ssigma)}}{\epsilon^2}.
\]

The first-order approximation of $\PPhi(\ssigma)$ as $\epsilon$ goes to zero is 
\[
 \PPhi (\ssigma) \approx 1 - \epsilon \sum_i \inner{}{\ttheta, \ssigma(i) \vv_i}
= 1 - \epsilon \inner{}{\ttheta, \ss}.
\]
So, 
$\expec{\hh}{ \PPhi(\ssigma)^2  }$ approaches 1 as $\epsilon$ approaches $0$ and
\[
\frac{\varh{\hh}{\PPhi(\ssigma)}}{\epsilon^2}
\rightarrow 
\frac{\varh{\hh}{\epsilon \inner{}{\ttheta, \ssigma}}}{\epsilon^2}
=
\varh{\hh}{\inner{}{\ttheta, \ss}},
\]
because variance is unchanged by addition of constants.

We conclude that 
\begin{align*}
\norm{\ttheta}^2 & \geq \varh{\hh}{\inner{}{\ttheta, \ss}}.\qedhere
\end{align*}

\end{proof}

\subsection{1-Subgaussian Concentration}

In our proof that $\ss$ is 1-subgaussian, we will need to integrate the result of \Cref{lem:cov} over a path of external fields.
Our proof will make use of the logarithm of the partition function, which we define to be the function 
\[
F (\hh) \defeq \ln \calZ(\hh).
\]

The following facts about $F(\hh)$ are well known.
See, for example, \cite[Lemma 12]{CLV23}.
\begin{claim}\label{clm:F}
The gradient and Hessian of $F$ are given by
\[
\nabla_{\hh} F(\hh) = \expec{\hh}{\ssigma}
\qquad \text{and} \qquad 
\nabla^2_{\hh} F(\hh) = \covh{\hh}{\ssigma},
\]
where 
\[ 
\covh{\hh}{\ssigma} = \expec{\hh}{ \ssigma \ssigma^T } - 
  \left( \expec{\hh} \ssigma \right)   \left( \expec{\hh} \ssigma \right)^T
\]
is the covariance matrix of $\ssigma$.
\end{claim}
\begin{proof}
To compute the gradient, consider the $i$th coordinate.
\begin{align*}
    \frac{\partial}{\partial \hh(i)} F(\hh)
    & = 
    \frac{\partial}{\partial \hh(i)} \ln Z(\hh)
    \\
    & = \frac{ \frac{\partial}{\partial \hh(i)} Z(\hh)}{Z(\hh)}
    \\
    & = \frac{ \sum_{\ssigma} \frac{\partial}{\partial \hh(i)} w_{\hh}(\ssigma)}{Z(\hh)}
    \\
    & = \frac{ \sum_{\ssigma} \frac{\partial}{\partial \hh(i)} \exp(\inner{}{\hh, \ssigma}) w(\ssigma)}{Z(\hh)}
    \\
    & = \frac{ \sum_{\ssigma}  \ssigma(i) \exp(\inner{}{\hh, \ssigma}) w(\ssigma)}{Z(\hh)}    
    \\ & = \expec{\hh}{\ssigma(i)}.
\end{align*}
We can similarly show that 
\[
    \frac{\partial^2}{\partial \hh(i) \partial \hh(j)} F(\hh)
     = 
     \expec{\hh}{\ssigma(i) \ssigma(j)} - \expec{\hh}{\ssigma(i)} \expec{\hh}{\ssigma(j)}.
\]
\end{proof}

We now prove a strengthening of Theorem 1 that allows for an external field $\hh$.
Theorem 1 is the special case with $\hh = \zzero$.
For that $\hh$, $\expec{\zzero}{\ssigma} = \zzero$,
  which simplifies the proof slightly.

\begin{theorem}\label{thm:subgaussianh}
Assume that the vectors $\vv_1, \ldots, \vv_n$ all have Euclidean at most 1 and that no three are parallel unit vectors.
Let $\hh \in \Reals{n}$ and
  let $\ssigma$ be chosen with probability proportional to 
  $w_{\hh}(\ssigma)$.
  Let $\ss = \sum_i \ssigma(i) \vv_i$.
  Then, $\ss$ is a 1-subgaussian random vector.
  That is, for all vectors $\ttheta$, 
  \[
  \expec{\hh}{\exp (\inner{}{\ttheta, \ss - \expec{\hh}{\ss}})}
  \leq \exp(\norm{\ttheta}^2 / 2).  
  \]
\end{theorem}

\begin{proof}
Let $\ttheta$ be a vector in $\Reals{d}$.
We will show 
\[
\ln \left( \expec{\hh}{\exp (\inner{}{\ttheta, \ss - \expec{\hh}{\ss}}) }\right)
\leq \norm{\ttheta}^2 / 2.
\]
To turn this into an expression in $\ssigma$, let $\VV$ be the matrix whose columns are $\vv_1, \ldots, \vv_n$ so that $\ss = \VV \ssigma$.
Define 
\[
\pphi = \VV^T \ttheta
\]
so that 
\[
\inner{}{\ttheta, \ss} = \inner{}{\pphi, \ssigma}.
\]
The connection to the partition functions is 
\begin{align*}
\expec{\hh}{\exp (\inner{}{\ttheta, \ss}) } 
& = 
\expec{\hh}{\exp (\inner{}{\pphi, \ssigma})}
\\
& = 
\frac{\calZ (\pphi + \hh)}{\calZ (\hh)},
\end{align*}
and so 
\[
\ln \left( \expec{\hh}{\exp (\inner{}{\ttheta, \ss - \expec{\hh}{\ss}}) }\right)
 = F(\pphi + \hh) - F(\hh) - 
 \inner{}{\pphi, \expec{\hh}{\ssigma}}.
\]
This reduces our problem to that of showing 
\[
 F(\pphi + \hh) - F(\hh)
\leq \norm{\ttheta}^2 / 2 + \inner{}{\ttheta, \expec{\hh}{\ss}} .
\]
We will do this by integrating on a path from $\hh$ to $\pphi + \hh$.
For a real variable $t \in [0,1]$, define 
\[
f(t) \defeq F(t \pphi + \hh).
\]
As $f(t)$ is twice continuously differentiable,
\[
f(1) - f(0) = \int_{t = 0}^{1} f'(t) dt 
= f'(0) + \int_{t=0}^{1} (1-t) f''(t) dt.
\]

By the chain rule, $f'(t) = \inner{}{\pphi, \nabla F (t \pphi + \hh)}$.
So \Cref{clm:F} implies $f'(0) = \inner{}{\pphi, \expec{\hh} \ssigma} = \inner{}{\ttheta, \expec{\hh}{\ss}}$.
The chain rule also gives
\begin{align*}
 f''(t) & = \pphi^T \nabla^2 F(t \pphi + \hh) \pphi  
 \\
 & = \pphi^T \covh{t \pphi + \hh}{\ssigma} \pphi
 \qquad \text{(by \Cref{clm:F})}
 \\
 & = \ttheta^T \VV \covh{t \pphi + \hh}{\ssigma} \VV^T \ttheta
 \\
 & = \ttheta^T \covh{t \pphi + \hh}{\VV \ssigma}  \ttheta
 \\
  & = \ttheta^T \covh{t \pphi + \hh}{\ss}  \ttheta
 \\
 & \leq \norm{\ttheta}^2,
\end{align*}
 by Lemma~\ref{lem:cov}.
Plugging this upper bound back in to the expression for $f(1) - f(0)$ proves the theorem:
\[
f(1) - f(0)
= 
f'(0) + 
\int_{t=0}^{1} (1-t) f''(t) dt
\leq 
f'(0) + 
\int_{t=0}^{1} (1-t) \norm{\ttheta}^2 dt
=
\inner{}{\ttheta, \expec{\hh}{\ss}} + 
\norm{\ttheta}^2 / 2. \qedhere
\]
\end{proof}

\section{Glauber dynamics}\label{sec:Glauber}

Next we analyse the standard algorithm to sample from the Gibbs distribution $\mu_h$ induced by~\eqref{eqn:tilted-weight}, namely Glauber dynamics.
The key result is a Poincar\'e inequality,
which is also derived from \Cref{thm:AF}.

For $i\in[n]$, let $P_i$ be the matrix that,  when acting on probability distributions represented by row vectors, resamples the $i$th coordinate.
Namely, $P_i(\ssigma)$ keeps $\ssigma_{-i}$, the partial configuration other than $i$, and replaces the $i$th coordinate by drawing from the distribution $\mu_h$ conditioned on $\ssigma_{-i}$. 
When applied to column vectors $f : \setof{\pm 1}^n \rightarrow \Reals{}$, $P_i$ replaces $f(\ssigma)$ with a weighted average of $f(\ssigma)$ and $f(\tau)$ where $\tau$ is the $\setof{\pm 1}^n$ vector that differs from $\ssigma$ in coordinate $i$: $(P_i f)(\ssigma) = (w_{\hh}(\ssigma) f(\ssigma) + w_{\hh}(\tau) f(\tau)) / (w_{\hh}(\ssigma) + w_{\hh}(\tau))$.
Let 
\[
P \defeq \frac{1}{n} \sum_{i=1}^{n} P_i.
\]
Also, define
\begin{align*}
  D_i\defeq I-P_i.
\end{align*}

Let $\alpha_i = 1 - \norm{\vv_i}^2$ and $\alpha_{0} = \min_i \alpha_i$. We prove the following approximate tensorization of variance for distribution $\mu_h$, which is equivalent to the Poincar\'e inequality for its Glauber dynamics. We refer readers to~\cite{CLV21} for a detailed introduction.

\begin{lemma}\label{lem:var_leq}
For every $f:\setof{\pm 1}^n\to\=R$,
\begin{align}\label{eqn:poincare-all-field}
  \varh{\hh}{f ( \ssigma )}
    \le\frac{1}{\alpha_{0}}\sum_{i=1}^n\=E_h\left[((D_i f)(\ssigma))^2\right].
\end{align}
\end{lemma}

Our proof rests on a consequence of \Cref{thm:AF} which we state, prove, and then massage to 
prove \Cref{lem:var_leq}. 
For $\zz$ and $\ssigma$ in $\Reals{n}$, we define the function
\[
\PPhi( \zz, \ssigma) \defeq \prod_{i=1}^{n} (1 - \zz(i) \ssigma(i)).
\]

\begin{lemma}\label{lem:generalz}
Assume that $\norm{\vv_i}^2 \leq 1 - \alpha_{0}$ for each $i$.
Let $S$ be a finite set of vectors in $\Reals{n}$
  and  for each $\zz \in S$ let $c_{\zz}$ be a real number.
If all vectors in $S$ have norm at most $\sqrt{\alpha_{0}}$,
then 
\begin{equation}\label{eqn:generalz}
\varh{\hh}{ \sum_{\zz \in S} c_{\zz} \PPhi(\zz, \ssigma)  } 
\leq
\frac{1}{\alpha_{0}}
\expec{\hh}{  \norm{\sum_{\zz \in S} c_{\zz} \zz \PPhi(\zz, \ssigma) }^{2}
  },
\end{equation}
where the expectation and variance are taken over $\ssigma$ distributed according to $w_{\hh}$.
\end{lemma}

We prove \Cref{lem:generalz} by adding coordinates to the vectors $\vv_i$ and $\zz$ in a way we describe below, and then applying the following lemma to reason about them.
The difference between \Cref{lem:generalz} and \Cref{lem:generalu} is that
  \Cref{lem:generalz} considers functions defined by inner products with $\ssigma$, 
  while \Cref{lem:generalu} considers inner products with different vectors.

For vectors $\uu$ and $\ww_1, \ldots, \ww_n$ of the same dimension,
define the function
\[
\PPhi_{\uu}^{\ww} (\ssigma) \defeq \prod_{i=1}^{n} (1 - \ssigma(i) \inner{}{\uu, \ww_i}).
\]

\begin{lemma}\label{lem:generalu}
Let $\ww_1, \ldots, \ww_n$ be vectors of norm at most 1 such that for all $j \neq k$
\[
\inner{}{\ww_j, \ww_k} = \inner{}{\vv_j, \vv_k}.
\]
Let $U$ be a finite set of vectors.
For each $\uu \in U$, let
 $c_{\uu}$ be a real number.
If all the vectors in $U$ have norm at most 1, then
\begin{equation}\label{eqn:generalu}
\varh{\hh}{ \sum_{\uu \in U} c_{\uu} \PPhi^{\ww}_{\uu}( \ssigma)  } 
\leq
\expec{\hh}{  \norm{\sum_{\uu \in U} c_{\uu} \uu \PPhi^{\ww}_{\uu}( \ssigma) }^{2}
  },
\end{equation}
where the expectation and variance are taken over $\ssigma$ distributed according to $w_{\hh}$.
\end{lemma}
\begin{proof}

Let $\BB$ be the Lorentz matrix of the vectors 
  $\setof{\pm \ww_i}_{i \in [n]}$, $\zzero$, and 
  $\setof{\uu}_{\uu \in U}$. 
The vectors $\pm \ww_i$ are indexed by coordinates $i^+$ and $i^-$
  and $\zzero$ is indexed by coordinate $0$.
For each $\uu \in U$, we index the vector $\uu$ by a coordinate we call $\uu$\footnote{
This overloading of the same symbol for a vector and a coordinate is convenient, and we hope it does not create confusion.
}.
We also use the selection vectors $\xx_i$ defined in \eqref{eqn:h_selection},
  and let $\xx_{0}$ be the elementary unit vector in direction $0$.
And, we introduce just one more selection vector $\xx_{-1}$ defined by 
\[
\xx_{-1}(\uu) = c_{\uu}, \quad \text{for $u \in U$.}
\]

For selection vectors $\aa$ and $\bb$, define 
\[
\calY(\aa, \bb) \defeq \calV_{\BB} (\aa, \bb, \xx_1, \ldots, \xx_n).
\]
As in the proof of \Cref{lem:cov}, we apply \Cref{thm:AF} to show
\begin{equation}\label{eqn:general1}
  \calY(\xx_0, \xx_{-1})^2 \geq \calY(\xx_0, \xx_0) \calY(\xx_{-1}, \xx_{-1}).
\end{equation}
The assumption $\inner{}{\ww_j, \ww_k} = \inner{}{\vv_j, \vv_k}$ guarantees that 
\begin{align*}
 \calY(\xx_0, \xx_0) 
 & = 
 \sum_{\ssigma \in \setof{\pm 1}^n}
 \exp(\inner{}{\hh, \ssigma})
 \prod_{\setof{i,j} \in \binom{[n]}{2}} (1 - \ssigma(i) \ssigma(j) \inner{}{\ww_i, \ww_j } )
\\ 
& = 
 \sum_{\ssigma \in \setof{\pm 1}^n}
 \exp(\inner{}{\hh, \ssigma})
 \prod_{\setof{i,j} \in \binom{[n]}{2}} (1 - \ssigma(i) \ssigma(j) \inner{}{\vv_i, \vv_j } )
\\
& = 
 \sum_{\ssigma \in \setof{\pm 1}^n}
 w_{\hh}(\ssigma)
\\
 & = 
 \calZ (\hh).
\end{align*}
We may similarly compute
\begin{align*}
 \calY(\xx_0, \xx_{-1}) 
 & = \sum_{\uu \in U} c_{\uu} \sum_{\ssigma \in \setof{\pm 1}^n}   w_{\hh}(\ssigma) \prod_{i=1}^n (1 -  \inner{}{\uu, \ssigma (i) \ww_i})
 \\
 & = \calZ(\hh) \expec{}{\sum_{\uu \in U} c_{\uu} \PPhi^{\ww}_{\uu} ( \ssigma)}
 \\
\calY(\xx_{-1}, \xx_{-1}) 
  & = \calZ(\hh) \expec{}{\sum_{\uu, \uuhat \in U} c_{\uu} c_{\uuhat} (1 - \inner{}{\uu, \uuhat})
    \PPhi^{\ww}_{\uu} ( \ssigma) \PPhi^{\ww}_{\uuhat}( \ssigma)}
    \\
    & = 
    \calZ(\hh)  \Expec{}{\sum_{\uu \in U} c_{\uu} \PPhi^{\ww}_{\uu} ( \ssigma) }^2
    - 
    \calZ(\hh) \expec{}{\sum_{\uu, \uuhat \in U} c_{\uu} c_{\uuhat} \inner{}{\uu, \uuhat}
    \PPhi^{\ww}_{\uu} ( \ssigma) \PPhi^{\ww}_{\uuhat}( \ssigma)}
    \\
    & = 
    \calZ(\hh)  \Expec{}{\sum_{\uu \in U} c_{\uu} \PPhi^{\ww}_{\uu} ( \ssigma) }^2
    - 
    \calZ(\hh) \expec{}{\norm{\sum_{\uu \in U} c_{\uu} \uu 
    \PPhi^{\ww}_{\uu} ( \ssigma)}^2 }    .
\end{align*}
Plugging these terms into \eqref{eqn:general1}, canceling  $\calZ(\hh)^2$, 
  and rearranging terms gives 
\begin{align*}
    \expec{}{\norm{\sum_{\uu \in U} c_{\uu} \uu 
    \PPhi_{\uu} ( \ssigma)}^2 } 
    & \geq 
    \Expec{}{\sum_{\uu \in U} c_{\uu} \PPhi^{\ww}_{\uu} ( \ssigma) }^2 
    - 
    \left( \expec{}{\sum_{\uu \in U} c_{\uu} \PPhi^{\ww}_{\uu} ( \ssigma)} \right)^2 
    \\
    & = 
    \varh{\hh}{\sum_{\uu \in U} c_{\uu} \PPhi^{\ww}_{\uu} ( \ssigma)}.
\end{align*}
\end{proof}

\begin{proof}[Proof of \Cref{lem:generalz}]
We construct $\ww_1, \ldots, \ww_n$ from $\vv_1, \ldots, \vv_n$ by setting
  \[
  \ww_i = \vv_i \oplus \sqrt{\alpha_0} \ee_i ,
  \]
where $\ee_i$ is the elementary unit vector in $\Reals{n}$ in direction $i$, 
  and $\oplus$ indicates concatenation of vectors. 
For $i \neq j$, $\inner{}{\ee_i, \ee_j} = 0$ and so 
  $\inner{}{\ww_i, \ww_j} = \inner{}{\vv_i, \vv_j}.$
By our assumption that $\norm{\vv_i}^2 \leq 1 - \alpha_0$, we can conclude that 
  $\norm{\ww_i}^2 = \norm{\vv_i}^2 + \alpha_0 \leq 1$.
For each $\zz \in S$, define
\[
\zzhat \defeq \zzero_d \oplus \zz  / \sqrt{\alpha_0},
\]
and
\[
U = \setof{\zzhat : \zz \in S}.
\]
Note that each $\zzhat$ 
 consists of $d$ zeros followed by a rescaling of $\zz$.
For each $\zz$ we naturally set $c_{\zzhat} = c_{\zz}$.
As $\norm{\zz} \leq \sqrt{\alpha_0}$, $\norm{\zzhat} \leq 1$.
We have established the conditions that allow us to apply \Cref{lem:generalu}.

To see that the left-hand sides of \eqref{eqn:generalu} and \eqref{eqn:generalz} are equal, note that our constructed vectors satisfy
\[
 \inner{}{\zzhat, \ww_i} = \zz(i).
\]
So, 
\[
\PPhi^{\ww}_{\zzhat}(\ssigma) 
= 
\prod_{i=1}^{n} (1 - \ssigma(i) \inner{}{\zzhat, \ww_i})
= 
\prod_{i=1}^{n} (1 - \ssigma(i) \zz (i))
=
\PPhi(\zz, \ssigma).
\]
The right-hand sides are equal because
\begin{align*}
\norm{\sum_{\zzhat \in U} c_{\zzhat} \zzhat \PPhi^{\ww}_{\zzhat}( \ssigma) }^{2}
& = 
\frac{1}{\alpha_0} \norm{\sum_{\zz \in S} c_{\zz} (\zzero_d \oplus \zz) \PPhi^{\ww}_{\zzhat}( \ssigma) }^{2}
\\ & = 
\frac{1}{\alpha_0} \norm{\zzero_d \oplus \sum_{\zz \in S} c_{\zz} \zz
  \PPhi( \zz, \ssigma) }^{2}
\\ & =
\frac{1}{\alpha_0} \norm{\sum_{\zz \in S} c_{\zz} \zz \PPhi(\zz, \ssigma) }^{2}. \qedhere
\end{align*}
\end{proof}

To apply \Cref{lem:generalz} to prove \Cref{lem:var_leq}, we need to express every function $f(\ssigma)$
  as a sum of the form $\sum_{\zz \in S} c_{\zz} \PPhi(\zz,  \ssigma),$
  where each $\zz$ has norm at most $\sqrt{\alpha_0}$. 
The difficulty is that we need for the right-hand sides of the inequalities established in these lemmas to line up as well.
If we were to ignore the right-hand sides and the norm restriction, we could 
  take $S = \setof{\pm 1}^n$ and for each $\zz \in S$
set
\[
c_{\zz} = 2^{-n} f(-\zz).
\]
To see what this would provide, note that 
 for $\zz , \ssigma \in \setof{\pm 1}^n$
\begin{equation}\label{eqn:phi_minus_z}
\PPhi (\zz, \ssigma) = 
\begin{cases} 2^n & \text{if $\ssigma = -\zz$} \\
0 & \textit{otherwise}.
\end{cases}
\end{equation} 
So we would have
\[
\sum_{\zz \in \setof{\pm 1}^n}
  c_{\zz} \PPhi(\zz, \ssigma)
  = 
  \sum_{\zz \in \setof{\pm 1}^n}
   2^n \oone_{\ssigma = -\zz} 2^{-n} f(-\zz) 
  = 
  f(\ssigma).
\]

We fix the defects in this construction in two phases.
First, we employ a sum over $S = \setof{0, \pm 1}^n$ and give a concrete expression for the coefficients $c_{\zz}$.
We then show that this can be replaced by a sum over $S = \setof{0, \pm \delta}^n$ for any $\delta > 0$.

\begin{lemma}\label{lem:zpm1}
Let $S = \setof{0, \pm 1}^n$.
For every function $f :\setof{\pm 1}^n \rightarrow \Reals{}$, there exist real coefficients $c_{\zz}$ for $\zz \in S$ so that 
\begin{equation}\label{eqn:zpm1_lhs}
 \sum_{\zz \in \setof{0, \pm 1}^n}
  c_{\zz} \PPhi(\zz, \ssigma)  = f(\ssigma),
\end{equation}
and for $i \in [n]$
\begin{equation}\label{eqn:zpm1}
 \sum_{\zz \in \setof{0, \pm 1}^n}
  c_{\zz} \zz(i) \PPhi(\zz, \ssigma)  
  = 
  - \ssigma(i) (D_i f) (\ssigma).
\end{equation}
\end{lemma}
\begin{proof}
Let $f_i = P_i f$, so $D_i f = f - f_i$.
Also define 
\[
  g = f - \sum_{i=1}^n f_i, \qquad \text{so } f = g + \sum_{i=1}^n f_i.
\]
Let $S_0 = \{\pm 1\}^n$. 
Equation \eqref{eqn:phi_minus_z} tells us that 
  the functions $\{\Phi(\zz,\cdot): \zz \in S_0\}$ are a basis of all functions on $\setof{\pm 1}^n$ .
Similarly, let $S_i = \{\zz : \zz(i) = 0 \text{ and }\zz(j) = \pm 1 \text{ for } j \neq i\}$.
The functions $\{\Phi(\zz,\cdot) : \zz \in S_i\}$ form a basis for all functions
  on $\setof{\pm 1}^n$ that do not depend on their $i$th coordinate.
As $f_i$ does not depend on its $i$th coordinate, there exist coefficients $c_{i,\zz}$ such that for all $\ssigma \in \{\pm 1\}^n$,
\begin{align*}
  g(\ssigma) = \sum_{\zz \in S_0} c_{0,\zz} \Phi(\zz,\ssigma) \qquad \text{and} \qquad f_i(\ssigma) = \sum_{\zz \in S_i} c_{i,\zz} \Phi(\zz,\ssigma).
\end{align*}
This implies
\begin{align*}
  f(\ssigma) = \sum_{i=0}^n \sum_{\zz \in S_i} c_{i,\zz} \Phi(\zz,\ssigma).
\end{align*}
Furthermore, by the identity $\zz(j) \Phi(\zz,\ssigma) = -\ssigma(j) \Phi(\zz,\ssigma)$ which holds when $\zz(j) \in \{\pm 1\}$ and $\zz(j) \Phi(\zz,\ssigma) = 0$  when $\zz(j) = 0$, it holds
\begin{align*}
  \sum_{i=0}^n \sum_{\zz \in S_i} c_{i,\zz} \zz(j) \Phi(\zz,\ssigma) &= - \sum_{i=0}^n \sum_{\zz \in S_i} c_{i,\zz} \ssigma(j) \Phi(\zz,\sigma) \oone_{i \neq j}\\
  & = - \ssigma(j) \left(g(\ssigma) + \sum_{i \not \in \setof{0, j}} f_i(\ssigma)\right)\\
  & = - \ssigma(j) (D_j f)(\ssigma). 
\end{align*}
For $\zz \in S_i$ for some $i$, set $c_{\zz} = c_{i, \zz}$.
This definition is valid because the sets $S_i$ are disjoint.
Set $c_{\zz}$ to zero for all other $\zz$.
\end{proof}

Before we can apply Lemma~\ref{lem:generalz}, we need to replace the set of evaluations points $\setof{0, \pm 1}^n$ with $\setof{0, \pm \delta}^n$ for $\delta$ small enough that
$\delta \sqrt{n} \leq \alpha_0$. 
This is easy, and we can do it for any $\delta > 0$.
Rather than deriving the formula for the coefficients, we just prove one exists.

To start, note that $\PPhi(\zz, \ssigma)$ and $\zz(i) \PPhi(\zz, \ssigma)$ are polynomials in $\zz$ in which every variable has degree at most 2.
Let $\calP^2_{\zz}$ be the set of such polynomials. 
We use the following consequence of \Cref{lem:zpm1}.

\begin{corollary}\label{cor:linear}
There exists a linear function $L$ mapping $\calP^2_{\zz}$ to the reals such that
\begin{equation}\label{eqn:zpmL_lhs}
 L \left[ \PPhi(\zz, \ssigma) \right]  = f(\ssigma),
\end{equation}
and for $i \in [n]$
\begin{equation}\label{eqn:zpmL_rhs}
  L \left[ \zz(i) \PPhi(\zz, \ssigma) \right]
  = 
  - \ssigma(i) (D_i f) (\ssigma).
\end{equation}
\end{corollary}
\begin{proof}
The evaluation of a polynomial is a linear function of its coefficients, as is the sum of evaluations.
\end{proof}

\begin{lemma}\label{lem:zpm_delta}
For any $\delta > 0$, let $S = \setof{0, \pm \delta}^n$.
For every function $f :\setof{\pm 1}^n \rightarrow \Reals{}$, there exist real coefficients $c_{\zz}$ for $\zz \in S$ so that 
\begin{equation}\label{eqn:zpm_del_lhs}
 \sum_{\zz \in S}
  c_{\zz} \PPhi(\zz, \ssigma)  = f(\ssigma),
\end{equation}
and for $i \in [n]$
\begin{equation}\label{eqn:zpm_del_rhs}
 \sum_{\zz \in S}
  c_{\zz} \zz(i) \PPhi(\zz, \ssigma)  
  = 
  - \ssigma(i) (D_i f) (\ssigma).
\end{equation}
\end{lemma}
\begin{proof}
Lagrange interpolation provides for every $\yy \in S$ a polynomial 
$p_{\yy}(\zz) \in \calP^2_{\zz}$ so that 
\[
p_{\yy}(\zz) = \begin{cases}
    1 & \text{if $\yy = \zz$} \\
    0 & \text{for every other $\zz \in S$.}
\end{cases}
\]
Moreover, for every $q(\zz) \in \calP^2_{\zz}$,
\[
q(\zz) = \sum_{\yy \in S} q(\yy) p_{\yy}(\zz).
\]
So, for any linear function $L : \calP^2_{\zz} \rightarrow \Reals{}$
  and any polynomial $q(\zz) \in \calP^2_{\zz}$,
\[
L[q(\zz)]  
= L \left[ \sum_{\yy \in S} q(\yy) p_{\yy}(\zz) \right]
= \sum_{\yy \in S}  q(\yy) L[p_{\yy}(\zz)] .
\]
That is, we may take $c_{\yy} = L[p_{\yy}(\zz)]$.
\Cref{cor:linear} provides the desired linear function $L$.
\end{proof}

\begin{proof}[Proof of \Cref{lem:var_leq}]
Let $f$ be any function from $\setof{\pm 1}^n$ to $\Reals{}$.
Set $\delta = \alpha_0 / \sqrt{n}$ and $S = \setof{0, \pm \delta}^n$.
By \Cref{lem:zpm_delta}, there exist real coefficients $c_{\zz}$ for $\zz \in S$
  so that 
\begin{equation*}
 \sum_{\zz \in S}
  c_{\zz} \PPhi(\zz, \ssigma)  = f(\ssigma)
  \quad \text{and for $i \in [n]$} \quad
   \sum_{\zz \in S}
  c_{\zz} \zz(i) \PPhi(\zz, \ssigma)  
  = 
  - \ssigma(i) (D_i f) (\ssigma).
\end{equation*}
To apply \Cref{lem:generalz},
  note that 
\[
\expec{\hh}{  \norm{\sum_{\zz \in S} c_{\zz} \zz \PPhi(\zz, \ssigma) }^{2}
  }
= 
\expec{\hh}{ \sum_{i \in [n]} 
  \left( \sum_{\zz \in S} c_{\zz} \zz(i) \PPhi(\zz, \ssigma) \right)^{2}
  }
  = 
  \expec{\hh}{
  \sum_{i \in [n]} (- \ssigma(i) (D_i f) (\ssigma))^2
  }
  =
    \expec{\hh}{
  \sum_{i \in [n]} ( (D_i f) (\ssigma))^2,
  }
\]
as $\sigma(i) \in {\pm 1}$.
So, \Cref{lem:generalz}
  gives 
\[
\varh{\hh}{f(\ssigma)}
\leq 
\frac{1}{\alpha_0}
    \expec{\hh}{
  \sum_{i \in [n]} ( (D_i f) (\ssigma))^2
  }
  = 
  \frac{1}{\alpha_0}
    \sum_{i \in [n]} \expec{\hh}{
   ( (D_i f) (\ssigma))^2.
  }
\]
\qedhere

\end{proof}

Now we are ready to prove \Cref{thm:spectral-gap}.

\begin{proof}[Proof of \Cref{thm:spectral-gap}]
  We consider the tilted distribution~\eqref{eqn:tilted-weight}.
  It is well known that the spectral gap of the Glauber dynamics $P$ has a variational characterisation \cite[Lemma 13.7]{LP17}:
  \begin{align*}
    \operatorname{gap}(P)
      =\inf_{\varh{\hh}{f}>0}
        \frac{\langle f,(I-P)f\rangle_h}{\varh{\hh}{f}},
  \end{align*}
  where $\langle f,g \rangle_h = \expec{\hh}{[fg]}$.
  Since $I-P = \frac{1}{n}\sum_{i=1}^n D_i$ and $D_i=I-P_i$ is an orthogonal projection,
  \begin{align*}
    \langle f,(I-P)f\rangle_h
      =\frac1n\sum_{i=1}^n\=E_h\left[(D_i f)^2\right].
  \end{align*}
  So, \Cref{lem:var_leq} implies
  $\operatorname{gap}(P)\ge \alpha_0/n$.

  It remains to justify the stated mixing bound. 
  Since $P$ is irreducible, reversible, and has only non-negative eigenvalues, there is a well-known relationship between the spectral gap and the mixing time \cite[Theorem 12.4]{LP17}
  \begin{align*}
   t_{\mathrm{mix},h}(\varepsilon) \le \left\lceil \frac{1}{\operatorname{gap}(P)}\log\left(\frac{1}{\varepsilon\mu_{\hh,\min}}\right)\right\rceil,
  \end{align*}
  where $\mu_{\hh,\min}\defeq\min_{\ssigma} \mu_h(\ssigma)$.
  We bound $\mu_{\hh,\min}$ next.
  By Cauchy-Schwarz, $\abs{\inner{}{\vv_i,\vv_j}}\le1-\alpha_0$.
  Together with $e^{-\norm{\hh}_1}\le e^{\inner{}{\hh,\ssigma}}\le e^{\norm{\hh}_1}$,
  this implies
  \begin{align*}
    \mu_{\hh,\min} & \ge2^{-n}e^{-2\norm{\hh}_1}
        \left(\frac{\alpha_0}{2-\alpha_0}\right)^{n^2}. \qedhere
  \end{align*}
\end{proof}

We now prove~\Cref{thm:spectral-gap-general-ising} and~\Cref{cor:SK-spectral-gap}.
\begin{proof}[Proof of~\Cref{thm:spectral-gap-general-ising}]
  Let $\rho \defeq \lambda_{\max}(\TT)$. Since $\rho I_n - \TT \succeq 0$, we may choose vectors $\vv_1,\dots,\vv_n$ such that
  \begin{align*}
    \forall i \neq j, \quad \inner{}{\vv_i,\vv_j} = - \TT_{ij} \qquad \text{and} \qquad \norm{\vv_i}^2 = \rho < 1.
  \end{align*}
  With this choice of vectors $\vv_1,\dots,\vv_n$, the tilted distribution~\eqref{eqn:tilted-weight} coincides with the general Ising model~\eqref{eqn:general-Ising}.
  Applying~\Cref{thm:spectral-gap} with $\alpha_0 = 1 - \rho$ yields the desired spectral gap bound.
\end{proof}

\begin{proof}[Proof of~\Cref{cor:SK-spectral-gap}]
  Since $\abs{\tanh x - x} \le \abs{x}^3$, the difference between $\TT$ and $\JJ$ satisfies
  \begin{align}\label{eqn:TT-JJ}
    \expec{}{\norm{\TT-\JJ}_{\mathrm{op}}^2} \le \expec{}{\norm{\TT-\JJ}_{\mathrm{F}}^2} \le \sum_{i \neq j} \expec{}{\JJ_{ij}^6} = O_\beta(n^{-1}).
  \end{align}
  With high probability 
   $\lambda_{\max}(\JJ) \leq 2\beta + o_n(1)$ 
   (see \cite{BaiYin} or \cite[Theorem 2.1.22]{anderson2010introduction}).
  Together with~\eqref{eqn:TT-JJ}, this implies $\lambda_{\max}(\TT) = 2\beta + o_n(1)$ with high probability.
  The desired spectral gap bound then follows from~\Cref{thm:spectral-gap-general-ising}.
\end{proof}

\section{General rapid mixing results}\label{sec:general}

It is natural to wonder to what extent the machinery in the last section generalises to other spin systems.
We first generalise it to a complete graph with uniform interactions, and then see how it can be applied to more general spin systems.

Let $q\ge2$ and $n\ge2$ be integers, and let $\AA\in\=R_{\ge0}^{q\times q}$ be a symmetric matrix with at most one positive eigenvalue, and no all-zero rows.
For $i\in[n]$, let $\llambda_i\in\=R_{\ge0}^q$ be a vector of external fields.
Define the partition function
\begin{align}\label{eqn:q-spin-partition}
  \+Z_{\AA}(\llambda_1,\dots,\llambda_n)
  \defeq\sum_{\ssigma\in[q]^n}
    \prod_{i=1}^n\llambda_i(\ssigma_i)
    \prod_{1\le i<j\le n}\AA_{\ssigma_i\ssigma_j}.
\end{align}
We assume $\calZ_{\AA}(\llambda_1,\ldots,\llambda_n) > 0$.
Define the Gibbs distribution
\begin{align*}
  \mu_{\AA,\llambda}(\ssigma)
  \defeq\frac{1}{\+Z_{\AA}(\llambda_1,\dots,\llambda_n)}
    \prod_{i=1}^n\llambda_i(\ssigma_i)
    \prod_{1\le i<j\le n}\AA_{\ssigma_i\ssigma_j}.
\end{align*}
Let $P_i$ be the averaging operator on site $i$, i.e., for every $f : [q]^n \to \Reals{}$ and $\ssigma$ in the support of $\mu_{\AA,\llambda}$,
\begin{align}\label{eqn:defPi}
  (P_i f)(\ssigma) = \expec{\ttau \sim \mu_{\AA,\llambda}}{\left[f(\ttau) \mid \ttau_{-i} = \ssigma_{-i}\right]}.
\end{align}
The Glauber dynamics $P$ for distribution $\mu_{\AA,\llambda}$ is defined by
\[
P \defeq \frac{1}{n} \sum_{i=1}^{n} P_i.
\]
Also, define
\begin{align*}
  D_i\defeq I-P_i.
\end{align*}

The Perron-Frobenius theorem gives a decomposition
\begin{align}\label{eqn:q-spin-Lorentz-representation}
  \AA=\rrho\transpose{\rrho}-\RR\transpose{\RR},
\end{align}
where $\rrho\in\=R_{>0}^q$ and $\RR\in\=R^{q\times r}$ for some $r\ge0$.
For $a\in[q]$, suppose the $a$-th row of $\RR$ is $\transpose{\rr_a}$, and let
\begin{align}\label{eqn:def-ca}
  \vv_a\defeq\frac{\rr_a}{\rrho_a},
  \quad
  c_a\defeq1-\norm{\vv_a}^2=\frac{\AA_{aa}}{\rrho_a^2},
  \quad\text{and}\quad
  c_{\AA}\defeq\min_{a\in[q]}c_a.
\end{align}
The entries of $\AA$ can be written as
\begin{align*}
  \AA_{ab} = \rrho_a \rrho_b (1-\inner{}{\vv_a,\vv_b}) \qquad \text{for $a,b \in [q]$}.
\end{align*}

We obtain the following spectral gap bound.
\begin{theorem}\label{thm:q-spin-spectral-gap}
  If $c_{\AA} > 0$, the Glauber dynamics $P$ satisfies $\operatorname{gap}(P)\ge\frac{c_{\AA}}{n}$.
\end{theorem}

In particular, if every diagonal entry of $\AA$ is positive, then $c_{\AA}>0$ and the spectral gap is $\Omega_{\AA}(1/n)$.

The proof of~\Cref{thm:q-spin-spectral-gap} follows the same strategy as that of~\Cref{thm:spectral-gap}. We first introduce the analogue of~\Cref{lem:generalu}.

\begin{lemma}\label{lem:q-spin-generalu}
Let $\{\ww_i^a\}_{i \in [n], a \in [q]} \subseteq \Reals{d}$ be vectors of norm at most 1 such that
\[
\inner{}{\ww^a_j, \ww^b_k} = \inner{}{\vv_a, \vv_b} \qquad \text{ for all $j \neq k$ and $a, b \in [q]$.}
\]
For $\uu \in \Reals{d}$, define
\begin{align*}
  \PPhi_{\uu}^{\ww}(\ssigma) \defeq \prod_{i=1}^n (1-\inner{}{\uu,\ww_i^{\ssigma_i}}).
\end{align*}
Let $U$ be a finite set of vectors in $\Reals{d}$.
For each $\uu \in U$, let
 $c_{\uu}$ be a real number.
If all the vectors in $U$ have norm at most 1, then
\begin{equation}\label{eqn:q-spin-generalu}
\varh{\AA,\llambda}{ \sum_{\uu \in U} c_{\uu} \PPhi^{\ww}_{\uu}( \ssigma)  }
\leq
\expec{\AA,\llambda}{  \norm{\sum_{\uu \in U} c_{\uu} \uu \PPhi^{\ww}_{\uu}( \ssigma) }^{2}
  },
\end{equation}
where the expectation and variance are taken over $\ssigma$ distributed according to $\mu_{\AA,\llambda}$.
\end{lemma}
\begin{proof}
  Let $\BB$ be the Lorentz matrix of the vectors 
  $\setof{\ww^a_i}_{i \in [n], a\in [q]}$, $\zzero$, and 
  $\setof{\uu}_{\uu \in U}$. 
The vectors $\ww^a_i$ are indexed by coordinates $i^a$
and $\zzero$ is indexed by coordinate $0$.
For each $\uu \in U$, we index the vector $\uu$ by a coordinate we call $\uu$.
% \footnote{
% This overloading of the same symbol for a vector and a coordinate is convenient, and we hope it does not create confusion.
% }.
For each $i \in [n]$, we define the selection vector $\xx_i$ by
\begin{align*}
  \xx_i(i^a) = \llambda_i(a) \rrho_a^{n-1} \qquad \text{for all $a \in [q]$},
\end{align*}
and set it to zero on all other coordinates.
Let $\xx_{0}$ be the elementary unit vector in direction $0$.
And, we introduce just one more selection vector $\xx_{-1}$ defined by 
\[
\xx_{-1}(\uu) = c_{\uu}, \quad \text{for $\uu \in U$,}
\]
and set it to zero on all other coordinates.
For selection vectors $\aa$ and $\bb$, define 
\[
\calY(\aa, \bb) \defeq \calV_{\BB} (\aa, \bb, \xx_1, \ldots, \xx_n).
\]
We apply \Cref{thm:AF} to show
\begin{equation}\label{eqn:q-spin-general1}
  \calY(\xx_0, \xx_{-1})^2 \geq \calY(\xx_0, \xx_0) \calY(\xx_{-1}, \xx_{-1}).
\end{equation}
The assumption $\inner{}{\ww^a_j, \ww^b_k} = \inner{}{\vv_a, \vv_b}$ guarantees that 
\begin{align*}
  \calY(\xx_0,\xx_0) &= \calZ_{\AA}(\llambda_1,\ldots,\llambda_n)\\
  \calY(\xx_0,\xx_{-1}) &= \calZ_{\AA}(\llambda_1,\ldots,\llambda_n) \expec{\AA,\llambda}{\left[\sum_{\uu \in U} c_{\uu} \PPhi^{\ww}_{\uu}( \ssigma)\right]}\\
  \calY(\xx_{-1},\xx_{-1}) &= \calZ_{\AA}(\llambda_1,\ldots,\llambda_n) \expec{\AA,\llambda}{\left[\left(\sum_{\uu \in U} c_{\uu} \PPhi^{\ww}_{\uu}( \ssigma)  \right)^2 -  \norm{\sum_{\uu \in U} c_{\uu} \uu \PPhi^{\ww}_{\uu}( \ssigma) }^{2}\right]}.
\end{align*}
Plugging these terms into~\eqref{eqn:q-spin-general1} and rearranging the terms gives~\Cref{lem:q-spin-generalu}.
\end{proof}

To apply~\Cref{lem:q-spin-generalu}, we use the following interpolation lemma.
For $\zz \in\Reals{nq}$, define
\[
  \PPsi_{\zz}(\ssigma)\defeq\prod_{i=1}^n(1-\zz_{i,\ssigma_i}).
\]

\begin{lemma}\label{lem:q-spin-zpm_delta}
For any $\delta>0$, let $S=\setof{0,\pm\delta}^{nq}$.
For every $f:[q]^n\to\Reals{}$, there are real coefficients $(b_{\zz})_{\zz\in S}$
such that, for every $\ssigma$ in the support of $\mu_{\AA,\llambda}$,
\begin{align}
  \sum_{\zz\in S}b_{\zz}\PPsi_{\zz}(\ssigma)
    &=f(\ssigma),\label{eqn:q-spin-zpm_lhs}\\
  \sum_{\zz\in S}b_{\zz}\zz_{i,a}\PPsi_{\zz}(\ssigma)
    &=(D_i f)(\ssigma)
    \qquad\text{for all $i\in[n]$ and $a\in[q]$.}\label{eqn:q-spin-zpm_rhs}
\end{align}
\end{lemma}
\begin{proof}
Recall that $\calP^2_{\zz}$ is the set of polynomials $g(\zz)$ in which every variable has degree at most $2$.
For this proof, extend $P_i$ to an operator on all $f:[q]^n \rightarrow \Reals{}$ by the same conditional-expectation formula \eqref{eqn:defPi} whenever the conditioning event has positive probability, and set it to zero otherwise.
Thus $P_i f$ depends only on $\ssigma_{-i}$.
Viewing $\PPsi_{\zz}(\ssigma)$ as a function of $\zz$, for any fixed $\ssigma\in[q]^n$, $i\in[n]$, and $a \in [q]$, both $\PPsi_{\zz}(\ssigma)$ and $(1-\zz_{i,a}) \PPsi_{\zz}(\ssigma)$ are in $\calP^2_{\zz}$.
We show the existence of a linear functional $L$ mapping $\calP_{\zz}^2$ to the reals such that, as a polynomial in $\zz$, 
\begin{align*}
  L[\PPsi_{\zz}(\ssigma)]=f(\ssigma) \qquad \text{and}\qquad L[(1-\zz_{i,a})\PPsi_{\zz}(\ssigma)]=(P_i f)(\ssigma).
\end{align*}
Suppose there are real coefficients $\aalpha_{\ssigma}$ and $\bbeta_{\ssigma,i,a}$ such that
\begin{align}
  \sum_{\ssigma}\aalpha_{\ssigma} \PPsi_{\zz}(\ssigma) + \sum_{\ssigma,i,a} \bbeta_{\ssigma,i,a} (1-\zz_{i,a}) \PPsi_{\zz}(\ssigma) = 0.
\end{align}
Set $\yy_{i,a} = 1 - \zz_{i,a}$. By comparing the monomials in $\yy$ on both sides, we must have
\begin{align}\label{eqn:coef-ab}
  \aalpha_{\ssigma} = 0 \qquad \text{and} \qquad \bbeta_{\ssigma,i,a} + \bbeta_{\ttau,i,b} = 0,
\end{align}
whenever $\ttau_{-i}=\ssigma_{-i}$, $\ssigma_i=b$, and $\ttau_i=a$.
Indeed, the first family consists of distinct monomials of total degree $n$, whereas the second family has total degree $n+1$; repeated monomials in the second family arise only by exchanging $a$ and $\ssigma_i$.
When $a=\ssigma_i$, the corresponding monomial occurs only once, so $\bbeta_{\ssigma,i,a}=0$.
Since $P_i f$ depends only on $\ssigma_{-i}$, 
\begin{align*}
  (P_i f)(\ssigma) = (P_i f)(\ttau).
\end{align*}
Plugging into~\eqref{eqn:coef-ab} gives
\begin{align*}
  \sum_{\ssigma} \aalpha_{\ssigma} f(\ssigma) + \sum_{\ssigma,i,a} \bbeta_{\ssigma,i,a} (P_i f)(\ssigma) = 0.
\end{align*} 
Thus the prescribed values respect all linear relations and define a linear functional on the span of these polynomials, which extends to $\calP_{\zz}^2$.

Lagrange interpolation provides for every $\xx \in S$ a polynomial
$p_{\xx}(\zz) \in \calP^2_{\zz}$ so that
\[
p_{\xx}(\zz) = \begin{cases}
    1 & \text{if $\xx = \zz$} \\
    0 & \text{for every other $\zz \in S$.}
\end{cases}
\]
Moreover, for every $g(\zz) \in \calP^2_{\zz}$,
\[
g(\zz) = \sum_{\xx \in S} g(\xx) p_{\xx}(\zz).
\]
So, for any linear functional $L : \calP^2_{\zz} \rightarrow \Reals{}$
  and any polynomial $g(\zz) \in \calP^2_{\zz}$,
\[
L[g(\zz)] = \sum_{\xx \in S}  g(\xx) L[p_{\xx}(\zz)] .
\]
Take $b_{\xx} = L[p_{\xx}(\zz)]$ for the functional constructed above.
Applying the interpolation identity to $\PPsi_{\zz}(\ssigma)$ gives~\eqref{eqn:q-spin-zpm_lhs}.
For $\ssigma$ in the support of $\mu_{\AA,\llambda}$, we also have
\[
  L[\zz_{i,a}\PPsi_{\zz}(\ssigma)]
  =L[\PPsi_{\zz}(\ssigma)]-L[(1-\zz_{i,a})\PPsi_{\zz}(\ssigma)]
  =f(\ssigma)-(P_i f)(\ssigma)=(D_i f)(\ssigma),
\]
which gives~\eqref{eqn:q-spin-zpm_rhs}.
\end{proof}

\begin{proof}[Proof of \Cref{thm:q-spin-spectral-gap}]
For simplicity, let $c = c_{\AA} > 0$.
Fix $0 < \epsilon < c$ and, in $\Reals{r}\oplus\Reals{n}\oplus\Reals{nq}$, let
\begin{align}
  \ww_i^a = \vv_a \oplus \sqrt{c-\epsilon} \ee_i \oplus \sqrt{\epsilon} \ee_{i,a},
\end{align}
where $\ee_i$ and $\ee_{i,a}$ are standard basis vectors in $\Reals{n}$ and $\Reals{nq}$, respectively.
A simple calculation yields $\norm{\ww_i^a}^2 = \norm{\vv_a}^2 + c \le 1$.
For all $j \neq k$ and $a,b \in [q]$,
\begin{align*}
  \inner{}{\ww^a_j, \ww^b_k} = \inner{}{\vv_a, \vv_b} + (c-\epsilon) \inner{}{\ee_j,\ee_k} + \epsilon \inner{}{\ee_{j,a},\ee_{k,b}} = \inner{}{\vv_a,\vv_b}.
\end{align*}
Furthermore, for each $\zz \in \Reals{nq}$, we let
\begin{align*}
\uu(\zz)= \zzero_r \oplus \sum_i \frac{\zz_{i,1}}{\sqrt{c-\epsilon}}\ee_i \oplus \sum_{i,a}\frac{\zz_{i,a}-\zz_{i,1}}{\sqrt\epsilon}\ee_{i,a}.
\end{align*}
The inner product $\inner{}{\uu(\zz),\ww_{i}^a} = \zz_{i,1} + \zz_{i,a}-\zz_{i,1} = \zz_{i,a}$. So,
\begin{align*}
  \PPhi_{\uu(\zz)}^{\ww}(\ssigma) = \prod_{i=1}^n (1-\inner{}{\uu(\zz),\ww_i^{\ssigma_i}}) = \prod_{i=1}^n (1-\zz_{i,\ssigma_i}) = \PPsi_{\zz}(\ssigma).
\end{align*}
Since $\uu(\zz)$ is linear in $\zz$, choose $\delta>0$ sufficiently small that $\norm{\uu(\zz)}\le1$ for every $\zz\in S\defeq\{0,\pm\delta\}^{nq}$.
Applying~\Cref{lem:q-spin-generalu} to the vectors $\{\uu(\zz):\zz\in S\}$ gives, for any real coefficients $b_{\zz}$,
\begin{align}\label{eq:var-2nd-moment}
  \varh{\AA,\llambda}{ \sum_{\zz \in S} b_{\zz} \PPsi_{\zz}( \ssigma)  }
\leq
\expec{\AA,\llambda}{  \norm{\sum_{\zz \in S} b_{\zz} \uu(\zz) \PPsi_{\zz}( \ssigma) }^{2}}.
\end{align}
For any $f:[q]^n\to\Reals{}$, take the coefficients supplied by~\Cref{lem:q-spin-zpm_delta} for this grid. Then
\begin{align*}
  \varh{\AA,\llambda}{ \sum_{\zz \in S} b_{\zz} \PPsi_{\zz}( \ssigma)  } = \varh{\AA,\llambda}{f(\ssigma)},
\end{align*}
and
\begin{align*}
  \expec{\AA,\llambda}{  \norm{\sum_{\zz \in S} b_{\zz} \uu(\zz) \PPsi_{\zz}( \ssigma) }^{2}} = \frac{1}{c-\epsilon} \sum_{i=1}^n \expec{\AA,\llambda}{((D_i f)(\ssigma))^2}.
\end{align*}
Combining with~\eqref{eq:var-2nd-moment} and taking $\epsilon \downarrow 0$ gives
\begin{align*}
  \varh{\AA,\llambda}{f(\ssigma)} \le \frac{1}{c} \sum_{i=1}^n \expec{\AA,\llambda}{((D_i f)(\ssigma))^2}.
\end{align*} 
This establishes the desired spectral gap lower bound for $P$.
\end{proof}

Now let $G=(V,E)$ be a graph with $\abs{V}=n$ and adjacency matrix $\operatorname{Adj}(G)$,
and $\AA$ be an interaction matrix for $q$ spins.
For external fields $\llambda=(\llambda_i)_{i\in V}$ with $\llambda_i\in\=R_{\ge0}^q$, define
\begin{align*}
  \+Z_{G,\AA}(\llambda)
  \defeq\sum_{\ssigma\in[q]^V}
    \prod_{i\in V}\llambda_i(\ssigma_i)
    \prod_{ij\in E}\AA_{\ssigma_i\ssigma_j}.
\end{align*}
Let $\mu_{G,\AA,\llambda}$ be the corresponding Gibbs distribution and let $P$ be the Glauber dynamics for $\mu_{G,\AA,\llambda}$.

The Perron--Frobenius theorem implies that $\AA=\rrho \transpose{\rrho}+\AA'$ for some $\rrho\in\=R_{>0}^q$ and matrix $\AA'$.
After shifting the weight of $\rrho$ to the external fields, and decomposing the remainder into positive and negative parts, we may assume that $\AA=\oone\transpose{\oone}+\AA^+-\AA^-$ for some matrices $\AA^+,\AA^-\succeq 0$.
Let $\lambda_{\max}\defeq\lambda_{\max}(\operatorname{Adj}(G))$ and $\lambda_{\min}\defeq\lambda_{\min}(\operatorname{Adj}(G))$ be the maximum and minimum eigenvalues of the adjacency matrix of $G$.
Set $\BB\defeq\lambda_{\max}\AA^+-\lambda_{\min}\AA^-$, and $c_{G,\AA}\defeq1-\max_{a\in[q]}\BB_{aa}$.

\begin{corollary}\label{cor:general-graph}
  With the notation above, if $c_{G,\AA}>0$, then $\operatorname{gap}(P)\ge\frac{c_{G,\AA}}{n}$.
\end{corollary}
\begin{proof}
To reduce this model to the complete-graph setting of~\eqref{eqn:q-spin-partition}, we keep the same vertex set and enlarge the spin set from $[q]$ to $V\times[q]$.
By assigning proper external fields $\widehat{\llambda}$, we restrict vertex $i$ to spins of the form $(i,a)$, where $a\in[q]$.
To preserve the weight of every configuration, the new interaction matrix $\AAhat$ must satisfy, for distinct $i,j\in V$,
\begin{align}\label{eqn:A-hat}
  \AAhat_{(i,a),(j,b)} = \begin{cases}
  \AA_{ab}, & ij \in E,\\
  1, & i \neq j \text{ and } ij \not\in E.
  \end{cases}
\end{align}

The remaining entries, with $i=j$, do not affect these configuration weights and can be chosen freely.
We choose them so that $\AAhat$ satisfies the hypotheses of~\Cref{thm:q-spin-spectral-gap}; in particular, it must be non-negative entrywise and have at most one positive eigenvalue.

We claim $\AAhat \defeq \oone \transpose{\oone} - \GGamma$ satisfies all constraints, where the  block matrix $\GGamma$, indexed by $V\times[q]$, is defined by
  \begin{align*}
    \GGamma
    \defeq \II_n\otimes \BB
      -\operatorname{Adj}(G)\otimes(\AA^+-\AA^-).
  \end{align*}
  % We claim that by defining $\widehat{\AA} \defeq \oone \transpose{\oone} - \Gamma$. 
  A straightforward calculation yields~\eqref{eqn:A-hat}. By the definition of $\BB$,
  \begin{align*}
    \GGamma
    ={}&\bigl(\lambda_{\max}\II_n-\operatorname{Adj}(G)\bigr)\otimes \AA^++\bigl(\operatorname{Adj}(G)-\lambda_{\min}\II_n\bigr)\otimes \AA^-.
  \end{align*}
  Both factors involving $\operatorname{Adj}(G)$ are positive semidefinite.
  Since $\AA^+,\AA^-\succeq0$, both tensor-product terms are positive semidefinite, and hence $\GGamma\succeq0$ and $\AAhat$ has at most one positive eigenvalue.
  Finally, the non-negativity of $\AAhat$ follows from
  \begin{align*}
    \AAhat_{(i,a),(i,b)} = 1 - \BB_{ab} \ge 1 - \sqrt{\BB_{aa} \BB_{bb}} > 0,
  \end{align*}
  where the first inequality follows from $\BB \succeq 0$.

  It remains to compute the quantity $c_{\AAhat}$ in~\eqref{eqn:def-ca}.
  Since $\GGamma\succeq0$, write $\GGamma=\RR\transpose{\RR}$.
  Then $\AAhat=\oone\transpose{\oone}-\RR\transpose{\RR}$ is a representation of the form~\eqref{eqn:q-spin-Lorentz-representation} with $\rrho=\oone$.
  For this representation,~\eqref{eqn:def-ca} gives
  \begin{align*}
    c_{\AAhat}
      =\min_{(i,a)\in V\times[q]}\AAhat_{(i,a),(i,a)}
      =1-\max_{a\in[q]}\BB_{aa}
      =c_{G,\AA}>0.
  \end{align*}
  Therefore, by~\Cref{thm:q-spin-spectral-gap}, the Glauber dynamics $\widehat{P}$ for the distribution $\mu_{\AAhat,\widehat{\llambda}}$ has the spectral gap bound $\operatorname{gap}(\widehat{P}) \ge \frac{c_{G,\AA}}{n}$.
  Since the Glauber dynamics $\widehat{P}$ can be identified with the Glauber dynamics $P$ for $\mu_{G,\AA,\llambda}$, we have
  \begin{align}
  \operatorname{gap}(P) & = \operatorname{gap}(\widehat{P}) \ge \frac{c_{G,\AA}}{n}. \qedhere
  \end{align}
\end{proof}

What \Cref{cor:general-graph} is saying is that $\lambda_{\max}$ controls the allowed interaction strength of the ferromagnetic part of $\AA$, whereas $\lambda_{\min}$ controls the anti-ferromagnetic part.
% We will discuss more of this after \Cref{cor:AF-Ising-spectral-gap} in the context of the Ising model.

When $G$ is a complete graph, $\lambda_{\min}=-1$.
If, in addition, $\AA$ is anti-ferromagnetic, namely $\AA^+=0$,
then the condition of \Cref{cor:general-graph} recovers that of \Cref{thm:q-spin-spectral-gap}.
More generally, on line graphs, where $\lambda_{\min}\ge-2$,
\Cref{cor:general-graph} implies a $\Omega(1/n)$ spectral gap bound whenever $c_{\AA} > 1/2$.
Previously, it was known \cite{DHJM21,CLV24} that Glauber dynamics is rapid mixing for any anti-ferromagnetic $2$-spin system on line graphs with bounded degree.
In comparison, the strength of \Cref{cor:general-graph} is that it applies to any $q\ge 2$, does not require bounded degree, and allows arbitrary external fields.

For the anti-ferromagnetic $q$-state Potts model, let $0<b<1$ be the penalty for equal spins.
After rescaling, the interaction matrix $\AA$ can be written as
\[
  \AA=\oone\transpose{\oone}
  -\frac{q(1-b)}{q-1+b}\left(\II_q-\frac1q\oone\transpose{\oone}\right).
\]
Writing $\ell \defeq-\lambda_{\min}(\operatorname{Adj}(G))$, \Cref{cor:general-graph} therefore gives
\[
  \operatorname{gap}(P)\ge\frac1n\left(1-\ell\frac{(q-1)(1-b)}{q-1+b}\right)
\]
whenever the expression in parentheses is positive. This extends the parameter range and improves the spectral-gap lower bound in the recent result of Chen and Wang~\cite[Corollary~6.7]{CW26}.

\section*{Acknowledgments}

We thank Yitan Wang for many helpful discussions at the start of this research effort.

\section*{AI Methodology}
In 2021, we conjectured the signed sum $\sum_{i} \ssigma_i \vv_i$ induced by the Ising distribution~\eqref{eqn:weight} is $1$-subgaussian and that the Glauber dynamics on the distribution is rapidly mixing.
Black-box optimization techniques were used to provide evidence for these conjectures.
We posed these conjectures to ChatGPT-5.6 Sol Ultra, using a prompt like that 
used to resolve the Cycle Double-Cover Conjecture~\cite{OpenAI2026CDCPrompt}. 
The first interaction affirmed the first conjecture using the Alexandrov-Fenchel inequality as the key tool, and established an $n^{O(1/\alpha_0)}$ mixing time bound using the local-to-global argument in~\cite{AL20}.
We then asked ChatGPT-5.6 Sol Ultra to apply the key idea directly to improve the mixing bound, which led to~\Cref{thm:spectral-gap}.
Upon digesting the proof, the authors recognized that \Cref{thm:spectral-gap} provides mixing criteria in general settings described in~\Cref{sec:general}.
The authors have substantially refined the proofs, both by thinking and using ChatGPT-5.6 and ChatGPT-6.

\bibliographystyle{alpha}
\bibliography{ref}

\end{document}